\documentclass{easychair}

\usepackage{doc}

\usepackage[T1]{fontenc}
\usepackage{float}
\usepackage{booktabs}
\usepackage{amsmath}
\theoremstyle{definition}
\newtheorem{definition}{Definition}
\newtheorem{theorem}{Theorem}
\usepackage{BOONDOX-calo}
\usepackage{graphicx}
\usepackage{enumitem}
\usepackage{mdframed}
\usepackage{tcolorbox}
\usepackage{xcolor}
\usepackage{tikz}
\usetikzlibrary{trees}
\usetikzlibrary{shapes.geometric, arrows, positioning}
\tikzstyle{block} = [rectangle, draw, text centered, rounded corners, fill=cyan!20, text width=13em, minimum height=2em]
\tikzstyle{subblock} = [rectangle, draw, text centered, fill=white, text width=7em, minimum height=2em]
\tikzstyle{subblock2} = [rectangle, draw, text centered, fill=white, text width=7em, minimum height=2.7em]
\tikzstyle{subblockCyan} = [rectangle, draw, text centered, fill=cyan!20, text width=7em, minimum height=2em]
\tikzstyle{subblockCyan2} = [rectangle, draw, text centered, fill=cyan!20, text width=7em, minimum height=2.7em]
\usepackage{enumitem}
\usepackage{bussproofs}

\newcommand{\RightRuleLabel}[2]{%
  \RightLabel{%
    \begin{tabular}{@{}l@{}}
      \textsc{#1}\\[-.2ex]
      \footnotesize #2
    \end{tabular}%
  }%
}

\newcommand{\TightRightRuleLabel}[2]{%
  \RightLabel{%
    \hspace*{-0.8em}%
    \begin{tabular}{@{}>{\footnotesize\upshape}l@{}}%
      {\normalsize\textsc{#1}}\\[-.2ex]%
      #2%
    \end{tabular}%
  }%
}
\newcommand{\vcenterbox}[1]{%
  \begingroup
    \setbox0=\hbox{#1}%
    \parbox[c]{\wd0}{\box0}%
  \endgroup
}
\newcommand{\RightRightRuleLabel}[2]{%
  \RightLabel{%
   \hspace*{-0.8em}%
    \vcenterbox{\textsc{#1}}%
    \enspace
    {\footnotesize\upshape \vcenterbox{\shortstack[l]{#2}}}%
  }%
}

\newcommand{\TightRuleLabel}[1]{%
  \RightLabel{%
    \hspace*{-0.8em}%
    \textsc{#1}%
  }%
}

\newcommand{\lpDollar}{\texttt{\char`\$}}

\usepackage{amssymb}
\usepackage{mathtools}
\usepackage{array}
\usepackage{longtable}
\newcolumntype{M}[1]{>{\raggedright\arraybackslash}p{#1}}

\usepackage{listings}
\definecolor{lightgrey}{RGB}{240,240,240}

\lstdefinelanguage{Lambdapi}
{
  inputencoding=utf8,
  extendedchars=true,
  numbers=left, 
  numberstyle=\color{gray}, 
  numberstyle={},
  tabsize=2,
  basicstyle={\ttfamily\fontsize{9pt}{10pt}\upshape},
  escapeinside={(*@}{@*)},
  backgroundcolor=\color{lightgrey},
  morekeywords=[1]{abort,admit,admitted,apply,as,assert,assertnot,associative,assume,begin,builtin,commutative,compute,constant,debug,end,fail,flag,focus,generalize,have,in,induction,inductive,infix,injective,left,let,notation,off,on,opaque,open,prefix,print,private,proofterm,protected,prover,prover_timeout,quantifier,refine,reflexivity,require,rewrite,right,rule,sequential,simplify,solve,symbol,symmetry,type,TYPE,unif_rule,verbose,why3,with},
  sensitive=true,
  keywordstyle=[1]\color{blue},
  morekeywords=[2]{modifier,nameOfSymbol,typeOfSymbol,proofTerm,proofScript,declaredType,\$patternVariable,rightHandSide},
  keywordstyle=[2]\color{darkgray}\itshape,
  morecomment=[l]{//},
  morecomment=[n]{/*}{*/},
  commentstyle={\itshape\color{red}},
  string=[b]{"},
  stringstyle=\color{orange},
  showstringspaces=false,
  literate=
  {λ}{$\lambda$}1
  {↪}{$\hookrightarrow$}1
  {→}{$\rightarrow$}1
  {Π}{$\Pi$}1
  {≔}{$\coloneqq$}1
  {⊢}{$\vdash$}1
  {≡}{$\equiv$}1
  {��}{$\mathbb{B}$}1
  {��}{$\mathbb{L}$}1
  {ℕ}{$\mathbb{N}$}1
  {α}{$\alpha$}1
  {β}{$\beta$}1
  {η}{$\eta$}1
  {π}{$\pi$}1
  {τ}{$\tau$}1
  {ω}{$\omega$}1
  {∧}{$\wedge$}1
  {≤}{$\le$}1
  {≠}{$\neq$}1
  {∉}{$\notin$}1
  {×}{$\times$}1
  {⋅}{$\cdot$}1
  {⊤}{$\top$}1
  {⊥}{$\bot$}1
  {¬}{$\neg$}1
  {∨}{$\lor$}1
  {∀}{$\forall$}1
  {∃}{$\exists$}1
  {⇒}{$\Rightarrow$}1
  {⸬}{$\colon\colon$}1
  {□}{$\Box$}1
  {σ}{$\sigma$}1
  {⤳}{$\rightsquigarrow$}1
  {ι}{$\iota$}1
  {��}{$\mathbb{L}$}1
  {⇔}{$\Leftrightarrow$}1
  {₁}{{\textsubscript{1}}}1
  {ₑ}{{\textsubscript{e}}}1
  {ᵢ}{{\textsubscript{i}}}1
  {₂}{{\textsubscript{2}}}1
  {ₙ}{{\textsubscript{n}}}1
}

% ...

\newcommand{\lpTerm}[1]{\colorbox{lightgrey}{\texttt{\small{#1}}}}
\newcommand{\lpKeyword}[1]{\textcolor{blue}{\texttt{\small{#1}}}}
\newcommand{\sys}[1]{\textsf{#1}}
\newcommand{\EPrule}[1]{\textsc{(#1)}}

\DeclareMathOperator\nn{ff}
\DeclareMathOperator\pp{tt}

\usepackage{multirow}
\newcommand{\tablecat}[2]{%
  \multirow{#2}{*}{%
    \rotatebox[origin=c]{90}{\footnotesize\emph{\shortstack[c]{#1}}}%
  }%
}

\usepackage{hyperref}

\newcommand{\RuleDef}[1]{%
  \hypertarget{rule:#1}{\EPrule{#1}}%
}

\newcommand{\RuleRef}[1]{%
  \hyperlink{rule:#1}{\EPrule{#1}}%
}

\usepackage{comment}
\usepackage{ifthen}
\newboolean{showcomments}
\setboolean{showcomments}{false}

\newif\ifHALversion
\HALversiontrue

\newcommand{\melanie}[1]{%
  \ifthenelse{\boolean{showcomments}}%
    {\textcolor{cyan}{[Melanie: #1]}}%
    {}%
}

\newcommand{\melaniem}[1]{%
  \ifthenelse{\boolean{showcomments}}%
    {\marginpar{\textcolor{cyan}{[Melanie: #1]}}}%
    {}%
}

\newcommand{\alex}[1]{%
  \ifthenelse{\boolean{showcomments}}%
    {\textcolor{magenta}{[Alex: #1]}}%
    {}%
}

\newcommand{\alexm}[1]{%
  \ifthenelse{\boolean{showcomments}}%
    {\marginpar{\textcolor{magenta}{[Alex: #1]}}}%
    {}%
}

\newcommand{\fred}[1]{%
  \ifthenelse{\boolean{showcomments}}%
    {\textcolor{green}{[Frédéric: #1]}}%
    {}%
}

\newcommand{\fredm}[1]{%
  \ifthenelse{\boolean{showcomments}}%
    {\marginpar{\textcolor{green}{[Frédéric: #1]}}}%
    {}%
}

\title{Formal Verification of Proofs from Automated Theorem Provers for Higher-Order Logic}

\author{Melanie Taprogge\inst{1,2}\thanks{Based upon work from the action CA20111 EuroProofNet supported by COST (European Cooperation in Science and Technology).} 
\and Frédéric Blanqui\inst{1}
\and Alexander Steen\inst{2}}

\institute{Université Paris-Saclay, ENS Paris-Saclay, LMF, CNRS, INRIA, France 
\email{\{melanie.taprogge,frederic.blanqui\}@inria.fr}\and Institute of Mathematics and Computer Science, University of Greifswald, Greifswald, Germany
\email{alexander.steen@uni-greifswald.de}}

\authorrunning{Taprogge et al.}

\titlerunning{Formal Verification of Proofs from ATPs for HOL}

\begin{document}

\maketitle

\begin{abstract}
We identify common challenges and requirements for verifying proofs from automated theorem provers in the Dedukti logical framework and develop a general methodology for deriving encodings of calculus rules and proof steps, including clausification. 
We then apply this methodology to the EP calculus for higher-order logic and integrate it into the automated theorem prover \sys{Leo-III}. The resulting prototype reconstructs about 80\% of generated proof steps automatically 
and lays the foundation for making \sys{Leo-III} the first higher-order automated theorem prover in the Dedukti framework, while providing a basis for cross-system reuse. 
The implementation uncovered several bugs in Leo-III.
\end{abstract}



\section{Introduction and Related Work}\label{sec:intro}

Computer-assisted reasoning spans various application domains and has given rise to a diverse array of proof systems. Amongst these are \emph{automated theorem provers} (ATPs), systems that attempt to autonomously prove a conjecture based on given assumptions. 
While many ATPs are based on first-order logic (FOL), with prominent examples such as \sys{Vampire} \cite{VampireProver} and \sys{E} \cite{Eprover},
there is increasing interest in higher-order reasoning with ATP systems including \sys{Satallax}~\cite{DBLP:conf/cade/Brown12}, \sys{Leo-III}~\cite{LeoIIIthesis,SteenB21} and, more recently, \sys{Vampire} and \sys{E}.\footnote{Vampire and E were historically designed for reasoning in FOL. The most recent versions, however, also support reasoning in higher-order logic.} 
The proof certificates ATPs output (typically in the \sys{TSTP} syntax~\cite{TSTP}) are not \emph{formal} proofs in the sense of complete derivations in a fixed calculus, as they usually omit or compress rule applications and instantiations. In this work, we aim to enable the verification of proofs emitted by the higher-order ATP \sys{Leo-III} through an internal construction of such formal and independently verifiable proofs. 
This approach contrasts with checkers attempting to reprove the individual steps of a given proof certificate using a different ATP. This idea was pioneered by \sys{GDV}~\cite{Sut_GDV}; a more recent tool following similar principles is \sys{Nörgler}~\cite{noerglerPaper}.  
While the latter approach increases trustworthiness, it typically does not provide a formal guarantee of correctness, unless provers that output formal and verifiable proofs are used as trusted systems (implemented in \cite{DBLP:conf/flairs/SutcliffeBB25}). 
Moreover, the reproving process
might fail due to non-trivial proof steps.

\emph{Interactive theorem provers} (ITPs), also referred to as \emph{proof assistants}, such as \sys{Isabelle} \cite{DBLP:books/sp/NipkowPW02}, \sys{Rocq} \cite{CoqCoqCorrect}, the \sys{HOL} family \cite{DBLP:journals/jfp/Hutton94}, and \sys{Lean} \cite{DBLP:conf/cade/MouraKADR15} allow human-guided reasoning in a setting aimed at fully formal and certified proofs. 
Most ITPs integrate so-called \emph{hammers} that invoke external ATPs and SAT/SMT solvers to discharge some proof obligations after translating the goals and premises to the systems' input language  (e.g.~\cite{DBLP:conf/cade/Paulson10,DBLP:journals/jar/CzajkaK18,DBLP:conf/cpp/GauthierK15}). Crucially, ITPs do not blindly trust external results, but attempt to reconstruct proofs, or rely on verified proof checkers (e.g.~\cite{DBLP:conf/cav/EkiciMTKKRB17}). 

In the context of ITPs, there exist two notable examples of automated systems capable of HOL reasoning. 
\sys{Duper}~\cite{DBLP:conf/itp/CluneQBA24} is a prover in the \sys{Lean} ecosystem, and the aforementioned HOL ATP \sys{Satallax} can output \sys{Rocq} proofs~\cite{teucke2011translating}. 
While both systems can generate independently verifiable HOL proofs, the underlying ideas and mechanisms differ from our objective:

While \sys{Duper} also supports TPTP problems, it is primarily designed as native \sys{Lean} automation rather than a standalone prover. This is reflected in its system description in the 2023 iteration of the annual ATP competition CASC (CASC-29~\cite{sutcliffe2024cade}), where the authors describe it as being at an early stage and state that it is not expected to compete with mature ATPs.

\sys{Satallax} does not translate the proof trace found by the ATP to a formal \sys{Rocq} proof, but instead relies on the minimal unsatisfiable core found during proof search, based on which it performs a second, restricted search to reconstruct a higher-order tableau refutation. This refutation is then output as a \sys{Rocq} proof~\cite{teucke2011translating}. While the reconstruction procedure is theoretically complete, 
\cite{teucke2011translating} demonstrates that it can in practice introduce considerable overhead.

In contrast to these two projects, we aim to directly translate the proof traces of a mature, standalone prover (\sys{Leo-III}) to an independently verifiable format.

Translations between different ITPs are also important, as they enable cross-verification and the exchange of libraries of formal proofs. 
Within the \sys{HOL} family, \sys{OpenTheory} provides a shared interchange format and standard library, with external checkers~\cite{hurd2011opentheory,DBLP:journals/jlap/Abrahamsson20}. Beyond intra-\sys{HOL} exchange, direct translations target \sys{Isabelle}~\cite{DBLP:conf/cade/ObuaS06,DBLP:conf/itp/KaliszykK13}, \sys{Metamath}~\cite{DBLP:journals/jfrea/Carneiro16}, \sys{Nuprl}~\cite{DBLP:conf/tphol/NaumovSM01}, and \sys{Rocq}~\cite{DBLP:conf/itp/KellerW10}.

However, such direct translations between existing systems may require up to a quadratic number of pairwise translators, each labor-intensive to build and maintain. A more economical solution is to use a \emph{logical framework} -- a meta-logic expressive enough to represent the object logics of many systems and enable translations between them -- as a pivot language. The $\lambda\Pi$-calculus, also known as LF \cite{HarperHP87}, is a classical example: Using \emph{dependent types}, it can express higher-order logic and encode its proofs via the \emph{propositions-as-types} principle~\cite{howard1980formulae}, reducing proof checking to type checking. 
\sys{Twelf}~\cite{pfenning1999system} implements LF and has been used for cross-system translations~\cite{schurmann2006executable}. \sys{MMT}~\cite{rabe2013scalable} is foundation-independent
and has been used to encode higher-order logic and export the \sys{HOL-Light} library~\cite{kaliszyk2014towards}. 
Much of this work automatically encodes and verifies higher-order proofs across interactive provers. This differs from our objective, as it transfers precise proof objects where all information required for formal verification is already explicit. 
Although more rigorous proof formats have been proposed for ATPs (see, for example,~\cite{sctptp2025}), 
current systems typically produce incomplete, informal proof certificates where crucial details must be reconstructed to obtain fully checkable proofs.


\paragraph{The \sys{Dedukti} Framework.}

A logical framework that supports the verification of automated reasoning and interoperability between systems is the \sys{Dedukti} framework~\cite{deduktiLogicalFramework}. It extends LF with rewriting~\cite{dershowitz90chapter,terese03book}, yielding the $\lambda\Pi$-calculus modulo theory~\cite{systemU}, in which types are identified modulo user-defined rewrite rules. Like LF, it follows the proofs-as-terms convention, but -- unlike the aforementioned encodings and translations -- it enables more compact, \emph{shallow} encodings of various object logics.
There is an ongoing effort to integrate different systems in the framework: 
A number of tools that translate proof libraries from well-known proof assistants to \sys{Dedukti} have already been made available, such as \sys{Holide} for \sys{OpenTheory} \cite{Holide}, \sys{Krajono} for \sys{Matita}, \sys{Coqine} for \sys{Rocq} \cite{BoespflugB12}, \sys{Personoj} for \sys{PVS} \cite{hondet20types} and, more recently, \sys{hol2dk} for \sys{HOL-Light} \cite{blanqui24lpar}.
Lastly, several automated provers can emit \sys{Dedukti} certificates, like the SMT solver \sys{ArchSAT}~\cite{DBLP:phd/hal/Bury19} and the FOL-ATP \sys{ZenonModulo}~\cite{DelahayeDGHH13}
. An encoding of resolution/superposition proofs in $\lambda\Pi$-modulo~\cite{burel2013shallow} underlies \sys{iProverModulo}~\cite{iProver} and the \sys{Dedukti} output of \sys{Vampire} \cite{komel2025case}.

Several more recent projects build on \href{https://github.com/Deducteam/lambdapi}{\sys{Lambdapi}}, an interactive proof assistant and checker~\cite{Bal_rewritingEngine} that extends \sys{Dedukti} with several features including implicit arguments and \emph{tactics} (commands that perform complex proof-term constructions). 
This enables the encoding of proofs as \emph{proof-scripts}, which simplify proof generation by providing greater flexibility than explicit proof-terms, while still allowing \sys{Lambdapi} to produce fully formal proof-terms from given scripts. Although this shifts some reconstruction work to the checker, it has not been a bottleneck in ongoing projects and makes the generated certificates more modular and easier to inspect and debug.

There is ongoing work on translating SMT proofs in the \sys{Alethe} format~\cite{DBLP:journals/corr/abs-2107-02354} into \sys{Lambdapi}, as well as on translating proofs checked by \sys{Ethos}~\cite{ethos-rep} 
(see~\cite{DunneBurel2026SMTToLambdaPi}). 
\sys{Goéland} is a tableau-based FOL ATP with export to both \sys{Lambdapi} and \sys{Rocq}.

\paragraph{Contribution.}
Currently, automated systems generating formally verifiable proofs in the Dedukti framework are restricted to FOL. This work addresses that limitation by presenting a general methodology for encoding and checking both the proof calculus and concrete proofs of the HOL ATP \sys{Leo-III} in \sys{Lambdapi}. We provide reusable encodings for the rules of the EP calculus, including higher-order features such as extensionality, unification constraints, and clausification. As a practical contribution, we implement the approach as an open-source extension of \sys{Leo-III} that generates \sys{Lambdapi} proof-scripts for supported proof steps.

\section{Preliminaries}\label{sec:prelim}

In the following section, the theoretical foundations of HOL and the $\lambda \Pi$-calculus modulo theory are briefly recalled. More details can be found in \cite{andrews2013introduction,systemU}.

\subsection{Higher-Order Logic (HOL)}

%

Versions of HOL in use nowadays differ in their axiomatizations. HOL ATPs are generally based on \emph{extensional type theory} (ExTT)~\cite{AutomationHOL,DBLP:journals/jsyml/Henkin50}, which assumes functional and propositional extensionality. We adopt the version of ExTT with Hilbert choice in this work and use the terms HOL and ExTT interchangeably in the following.
The strategies demonstrated here are, however, 
modular and can be extended to different axiomatizations.

HOL is a typed logic based on Church’s simply typed $\lambda$-calculus~\cite{Church40}. Types are either \emph{base types} (sometimes also referred to as sorts), including $o$ (type of truth values) and $\iota$ (type of individuals), or \emph{function types} $T \rightarrow S$, for any types $T$ and $S$. 
Types are indicated using subscripts (e.g., $a_o$) 
or given explicitly as type assignment (e.g., $a : o$). They are omitted if clear from the context.

Terms and propositions are built from typed constants (e.g., connectives and quantifiers), typed variables, application, and abstraction. The choice of primitives is formalism-dependent but does not affect the results discussed here.

\subsection{\texorpdfstring{$\lambda\Pi$-calculus modulo theory}{lambda-Pi-calculus modulo theory}}
\label{subsec:lpm}

LF is an extension of the simply typed $\lambda$-calculus with
dependent types 
$\Pi x \colon \! T . \, S$, allowing the type $S$ to depend on a variable $x$ of type $T$, which coincides with $T\rightarrow S$ when $x$ does not occur in $S$. Extending the typing hierarchy of HOL with $Type$ (the type of types) and $Kind$ (the type of $Type$) allows for a unified syntax of LF \cite{DBLP:books/daglib/0032840}:

\begin{equation*}
    T,S = x \,|\, Type \,|\, Kind \,|\, \Pi x \colon \! T. \, S \,|\, \lambda x\colon \! T. \, S \,|\, T \, S
\end{equation*}

 
 
$\lambda\Pi$-calculus modulo theory extends this system with rewrite rules. These are pairs $(l,r)$ in $\beta$-normal form, denoted $l \hookrightarrow r$, expressing that any occurrence matching $l$ can be replaced with $r$. 
The relation $\rightarrow_{\beta \mathcal{R}}$ combines $\beta$-reduction with a set of rewrite rules $\mathcal{R}$~\footnote{$\rightarrow_{\beta \mathcal{R}}$ can be extended with $\eta$-conversion. In \sys{Lambdapi}, this is optional and enabled by a flag.}. The typing rules of the $\lambda \Pi$-calculus (see~\cite{systemU}) enforce well-formedness of terms. In particular, the domains of products and abstractions must be typable as $Type$, and $\rightarrow_{\beta \mathcal{R}}$ is incorporated in the conversion rule, identifying terms modulo these reductions. 

\subsection{Encoding HOL in \sys{Lambdapi}}\label{sec:encHOL}




A \emph{theory} of an object-logic in $\lambda\Pi$-calculus modulo theory is a set of type declarations and rewrite rules that enable the expression of the types, terms, and proofs of the object-logic. Using a shared theory 
for different deduction systems is advantageous 
as it allows us to prove desirable properties of encodings once and for all and facilitates the integration of encoded proofs derived by different tools.
A universal theory 
in \sys{Lambdapi} is given in the \href{https://github.com/Deducteam/lambdapi-stdlib}{\emph{standard library}} (based on \cite{systemU}). Different fragments of the theory can be selectively combined to encode various object-logics. User-defined extensions 
can also be introduced as needed. 
The encoding of HOL in the $\lambda\Pi$-calculus is well known, dating back to \cite{HarperHP87}. Here we illustrate the concrete \sys{Lambdapi} representation using examples. 
For readability, auxiliary commands (e.g., notation definitions) are omitted. 
See the \sys{Lambdapi} \href{https://lambdapi.readthedocs.io/}{manual} for a comprehensive guide to \sys{Lambdapi}. 
\ifHALversion
  The full HOL encoding and its formal translation are given in Appendix~\ref{ap:HOLenc}.
\else
  The full HOL encoding and its formal translation are given in the extended version of this paper~\cite{taprogge:hal-05735864}.
\fi




\paragraph{Types and constants}
are declared using \lpKeyword{symbol}, possibly preceded by properties such as \lpKeyword{constant}, which prohibits the symbol from being defined by rewrite rules. The name of the symbol is followed by \lpTerm{:} and its type:

\begin{lstlisting}
constant symbol Set : TYPE;
constant symbol o : Set;
injective symbol τ : Set → TYPE;
constant symbol Prop : TYPE;
rule τ o ↪ Prop;
constant symbol ⤳ : Set → Set → Set;
rule τ ($x ⤳ $y) ↪ τ $x → τ $y;
\end{lstlisting}

The constant \lpTerm{Set} is declared as a meta-level type (line 1) and serves as the foundation for encoding object-level types, such as \lpTerm{o} (line 2). 
\lpTerm{$\tau$} (line 3) interprets encoded HOL types as meta-level
types and is declared as \lpKeyword{injective}, meaning that $t
\equiv_{\beta\mathcal{R}} u$ whenever $\tau\,t
\equiv_{\beta\mathcal{R}} \tau\,u$\footnote{\sys{Lambdapi} does not
verify injectivity but, here, it can be easily proved by induction on
the size of $t$, once the confluence of
$\longrightarrow_{\beta\mathcal{R}}$ is established (see 
\ifHALversion
  Appendix~\ref{ap:HOLenc}).
\else
  \cite{taprogge:hal-05735864}).
\fi Injectivity is used in \sys{Lambdapi} unification
engine to solve convertibility constraints generated by type
checking.}. Function types are represented using \lpTerm{$\leadsto$}
(line 6), which is reduced to the meta-level function type constructor
by the rewrite rule declared in line 7 using the keyword
\lpKeyword{rule}. Variables prefixed by \lpTerm{\lpDollar}, such as
\lpTerm{\lpDollar x} and \lpTerm{\lpDollar y}, denote pattern
variables.



\paragraph{Encoding of terms.}

Examples of encoded object-logic symbols are shown below:

\begin{lstlisting}
constant symbol ⊥ : Prop;
constant symbol ⇒ : Prop → Prop → Prop;
symbol ¬ p ≔ p ⇒ ⊥;
constant symbol ∧ : Prop → Prop → Prop;
symbol = [t] : τ t → τ t → Prop;
constant symbol ∀ [t] : (τ t → Prop) → Prop;
symbol (*@$\varepsilon$@*) [a:Set] : (τ a → Prop) → τ a;
\end{lstlisting}

The symbols for the proposition for falsity (line 1), implication (line 2), and conjunction (line 4) follow standard type assignments. Negation (line 3) is defined with respect to implication and $\bot$. 
Dependent typing allows for polymorphic operators. Here, this is used to declare equality (line 5), universal quantification (line 6) and the choice operator (line 7).
Square brackets are used to mark the arguments as implicit, allowing the user to omit them when they can be inferred by \sys{Lambdapi}. 
The definitions of additional connectives are given in \cite{systemU}. 

\paragraph{Encoding of proofs.}

Proofs are encoded in $\lambda\Pi$-calculus modulo theory following the propositions-as-types principle, where types can represent propositions, and terms of these types serve as their proofs. Implementing this in \sys{Lambdapi} necessitates a few additional symbols, among them the following:

\begin{lstlisting}
injective symbol π : Prop → TYPE;
symbol ∧ₑ₁ [p q] : π (p ∧ q) → π p;
rule π ($p ⇒ $q) ↪ π $p → π $q;
rule π (∀(λ x, $f.[x])) ↪ Π x, π $f.[x];
\end{lstlisting}

The symbol \lpTerm{$\pi$} (line 1) is a mapping of propositions to
\lpKeyword{TYPE}, enabling the assignment of propositions-as-types. This is used to declare the rules of \emph{natural deduction} (ND) \cite{sep-natural-deduction}, as seen in the declaration of the elimination rule $\land_{e1}$ (line 2).
The Curry–Howard correspondence identifies implication and universal quantification in object terms with function types and dependent types in the types-as-terms encoding. This is implemented by the rewrite rules in lines 3-4. 

\paragraph{Axiomatization.}


To verify proofs, the foundational assumptions of the underlying logic are introduced as axioms. In the case of \sys{Leo-III}, this encompasses choice introduction as well as functional and propositional extensionality
. 
Excluded middle (\lpTerm{$em$:$\Pi$ p, $\pi$(p $\lor$ $\neg$ p)}) and double negation elimination (\lpTerm{$\neg\neg_e$:$\Pi$ x, $\pi$($\neg\neg$x) $\to$ $\pi$ x}) can then be proved from the above assumptions~\cite{Diaconescu1975Axiom}.

\begin{center}
\begin{lstlisting}
symbol (*@$\varepsilon$@*)ᵢ [a:Set] (p:τ a → Prop): π(∃ p) → π(p ((*@$\varepsilon$@*) p));
symbol funExt [a b](f g : τ(a ⤳ b)): (Π x, π(f x = g x)) → π(f = g);
symbol propExt x y: (π x → π y) → (π y → π x) → π (x = y);
\end{lstlisting}
\end{center}

\paragraph{Meta-properties.}
Since we add no new rewrite rules, the encodings we reuse from the \sys{Lambdapi} standard library inherit their meta-properties: correctness of the encoding (shown in~\cite{DBLP:conf/types/Grienenberger22}) and decidability of type checking (shown in~\cite{systemU}). We briefly recall the relevant properties and proofs in 
\ifHALversion
  Appendix~\ref{ap:HOLenc}.
\else
  \cite{taprogge:hal-05735864}.
\fi

\subsection{The proof calculus EP for higher-order logic}

EP, as proposed in~\cite{DBLP:phd/dnb/Benzmuller99,ganzinger1999extensional}, is an
extensional paramodulation calculus tailored to the efficient automation of HOL reasoning.

In the following, a literal is a signed equation $[s \simeq t]^\alpha$ with polarity $\alpha \in \{\pp,\nn\}$, and a clause is a disjunction of literals.
$[s]^\alpha$ is used as a shorthand notation for $[s \simeq \top]^\alpha$. 
Any free variables of a clause are implicitly universally quantified. 
For terms, $t\{s/x\}$ denotes the substitution of $s$ for the
variable $x$ in $t$. The subterm of $t$ at position $p$ is denoted by $t|_p$, and $t[s]_p$ is $t$ with the subterm at $p$ replaced by $s$. The application of a substitution $\sigma$ to a term $t$ and a clause $C$ is given by $t\sigma$ and $C\sigma$, respectively. The set of free variables of a term $t$ is denoted by $fv(t)$.
A central device in EP is the use of unification constraints~\cite{DBLP:phd/dnb/Kohlhase94,ganzinger1999extensional} to make rule applications conditional on the unifiability of two terms. Such constraints are negative equational literals of the form $[s \simeq t]^{\nn}$. They are introduced in the conclusions of several rules, where they record unification obligations, and are discharged by dedicated unification rules.


\paragraph{Primary inference rules.}

The three central rules of EP are \emph{paramodulation} \EPrule{Para}, \emph{primary substitution} \EPrule{Prim}, and \emph{factoring} \EPrule{Fac}. Paramodulation uses positive equational literals for conditional rewriting within clauses. Primary substitution introduces unification constraints that instantiate flexible head symbols using candidates from the set $\mathcal{GB}$ of \emph{general bindings}, which are used for the step-wise approximation of the structure of candidate terms, see~\cite{DBLP:journals/jsc/SnyderG89}. Factoring adds the constraints needed to unify two literals of a clause, enabling their contraction once the constraints are discharged.



\begin{minipage}{\textwidth}
\begin{centering}

\begin{prooftree}
  \AxiomC{$C \lor [s_T \simeq t_T]^\alpha$}
  \AxiomC{$D \lor [l_\nu \simeq r_\nu]^{\pp}$}
  \TightRightRuleLabel{\RuleDef{Para}}{($s|_\pi:\nu \ \wedge\ fv(s|_\pi)\subseteq fv(s)$)}
  \BinaryInfC{$[s[r]_\pi \simeq t]^\alpha \lor C \lor D \lor [s|_\pi \simeq l]^{\nn}$}
\end{prooftree}

\vspace*{-6mm}
\begin{prooftree}
      \AxiomC{$C \lor \big[\,H_{\tau}\,\overline{s^{\,i}}_{\tau_i}\,\big]^{\alpha}$}
      \AxiomC{$G \in \mathcal{GB}_{\tau}^{\{\neg,\vee\}\,\cup\,\{\Pi^{\nu},\,=^{\nu}\mid \nu\in\mathcal{T}\}}$}
      \TightRuleLabel{\RuleDef{Prim}}
      \BinaryInfC{$C \lor \big[\,H_{\tau}\,\overline{s^{\,i}}_{\tau_i}\,\big]^{\alpha}
                     \lor \big[\,H \simeq G\,\big]^{\nn}$}
\end{prooftree}

\end{centering}
\end{minipage}

\vspace*{1mm}
\begin{minipage}{\textwidth}
\begin{centering}

\begin{prooftree}
    \AxiomC{$C \lor [s_T \simeq t_T]^{\alpha} \lor [u_T \simeq v_T]^{\alpha}$}
    \TightRuleLabel{\RuleDef{Fac}}
    \UnaryInfC{$C \lor [s_T \simeq t_T]^{\alpha} \lor 
    [s_T \simeq u_T]^{\nn} \lor
    [t_T \simeq v_T]^{\nn}$}
\end{prooftree}

\end{centering}
\end{minipage}

\paragraph{Unification rules.}

EP complements the core rules with techniques designed to address challenges in HOL automation through the addition of dedicated calculus rules. The first of these challenges is \emph{unification}, which is undecidable in HOL. 
The unification constraints introduced above address this challenge and 
are handled by a dedicated group of unification rules implementing a version of Huet's pre-unification procedure~\cite{DBLP:journals/tcs/Huet75}. \EPrule{Triv} removes a constraint that has become trivially false after instantiation, while \EPrule{Bind} performs the substitution that realizes the constraint and drops it.
\EPrule{FlexRigid} and \EPrule{FlexFlex} introduce unification constraints binding quantified variables, 
and \EPrule{Decomp} 
decomposes unification constraints with identical heads into constraints on arguments.


\begin{minipage}{\textwidth}
\begin{centering}

    \begin{minipage}{0.8\textwidth}
        \begin{minipage}{0.4\linewidth}
            \begin{prooftree}
                \AxiomC{$C \lor [s_T = s_T]^{\nn}$}
                \TightRuleLabel{\RuleDef{Triv}}
                \UnaryInfC{$C$}
            \end{prooftree}
        \end{minipage}
        \hfill
        \begin{minipage}{0.5\linewidth}
            \begin{prooftree}
                \AxiomC{$C \lor [\mathcal{x}_T = s_T]^{\nn}$}
                \TightRightRuleLabel{\RuleDef{Bind}}{where $\mathcal{x}_T \notin fv(s)$}
                \UnaryInfC{$C\{s/\mathcal{x}\}$}
            \end{prooftree}
        \end{minipage}
    \end{minipage}
    
    \begin{prooftree}
      \AxiomC{$C \lor [c \, \overline{s^{i}} \, \simeq \, c \,\overline{t^{i}}]^{\nn}$}
      \TightRuleLabel{\RuleDef{Decomp}}
      \UnaryInfC{$C \lor [s^1 \simeq t^1]^{\nn} \lor ... \lor  [s^n \simeq t^n]^{\nn}$}
    \end{prooftree}

    \begin{prooftree}
      \AxiomC{$C \lor \big[\,X_{\overline{\mu} \to \nu}\,\overline{s^{\,i}} \simeq c_{\overline{\tau} \to \nu}\,\overline{t^{\,j}}\,\big]^{\nn}$}
      \AxiomC{$g_{\overline{\mu} \to \nu}\in \mathcal{GB}^{\{c\}}_{\overline{\mu} \to \nu}$}
      \TightRuleLabel{\RuleDef{FlexRigid}}
      \BinaryInfC{$C \lor \big[\,X_{\overline{\mu} \to \nu}\,\overline{s^{\,i}} \simeq c_{\overline{\tau} \to \nu}\,\overline{t^{\,j}}\,\big]^{\nn}
                     \lor \big[\,X \simeq g\,\big]^{\nn}$}
    \end{prooftree}
    
    \begin{prooftree}
      \AxiomC{$C \lor \big[\,X_{\overline{\mu} \to \nu}\,\overline{s^{\,i}} \simeq Y_{\overline{\tau} \to \nu}\,\overline{t^{\,j}}\,\big]^{\nn}$}
      \AxiomC{$g_{\overline{\mu} \to \nu}\in \mathcal{GB}^{\{h\}}_{\overline{\mu} \to \nu}$}
      \TightRightRuleLabel{\RuleDef{FlexFlex}}{where $h \in \Sigma$ is an \\appropriate constant}
      \BinaryInfC{$C \lor \big[\,X_{\overline{\mu}  \to \nu}\,\overline{s^{\,i}} \simeq Y_{\overline{\tau}  \to \nu}\,\overline{t^{\,j}}\,\big]^{\nn}
                     \lor \big[\,X \simeq g\,\big]^{\nn}$}
    \end{prooftree}
    
\vspace{0.2em}
    

\end{centering}
\end{minipage}

\paragraph{Extensionality rules.}


Extensionality is captured by 
\EPrule{PFE} and \EPrule{NFE} for functional extensionality, and \EPrule{PBE} and \EPrule{NBE} for propositional (Boolean) extensionality~\footnote{Note that the term \emph{extensionality} is commonly used for the reverse direction of the inference rules shown here, i.e., for deriving the equality of functions from their equality on every argument, and for deriving the equality of propositions from their logical equivalence. The names used here are, however, inherited from the \sys{Leo-III} project and are used for consistency with the existing literature.}~\cite{DBLP:cut-sim,AutomationHOL}.


\begin{minipage}{\textwidth}
\begin{centering}
\raggedright
\begin{minipage}{0.9\textwidth}  
        \begin{minipage}{0.45\linewidth}
        \centering
            \begin{prooftree}
                \AxiomC{$C \lor [s_o \simeq t_o]^{\pp}$}
                \TightRuleLabel{\RuleDef{PBE}}
                \UnaryInfC{$C \lor [s_o]^{\pp} \lor [t_o]^{\nn}$}
                \noLine
                \UnaryInfC{$C \lor [s_o]^{\nn} \lor [t_o]^{\pp}$}
                \end{prooftree}
            \end{minipage}
        \hfill
        \begin{minipage}{0.45\linewidth}
        \centering
            \begin{prooftree}
                \AxiomC{$C \lor [s_o \simeq t_o]^{\nn}$}
                \TightRuleLabel{\RuleDef{NBE}}
                \UnaryInfC{$C \lor [s_o]^{\pp} \lor [t_o]^{\pp}$}
                \noLine
                \UnaryInfC{$C \lor [s_o]^{\nn} \lor [t_o]^{\nn}$}
            \end{prooftree}
        \end{minipage}
\end{minipage}

\begin{minipage}{0.9\textwidth}  
        \begin{minipage}{0.45\linewidth}
        \centering
           \begin{prooftree}
                    \AxiomC{$C \lor [s_{T \rightarrow S} \simeq t_{T \rightarrow S}]^{\pp}$}
                    \TightRightRuleLabel{\RuleDef{PFE}}{$X_{T}$ is fresh}
                    \UnaryInfC{$C \lor [s\ X_T \simeq t\ X_T]^{\pp}$}
                    \singleLine
                \end{prooftree}
            \end{minipage}
        \hfill
        \begin{minipage}{0.45\linewidth}
        \centering
            \begin{prooftree}
                    \AxiomC{$C \lor [s_{T \rightarrow S} \simeq t_{T \rightarrow S}]^{\nn}$}
                    \TightRightRuleLabel{\RuleDef{NFE}}{$sk_T$ is a new\\Skolem term}
                    \UnaryInfC{$C \lor [s\ sk_T \simeq t\ sk_T]^{\nn}$}
                    \singleLine
                \end{prooftree}
        \end{minipage}
\end{minipage}

\end{centering}
\end{minipage}

\melaniem{briefly explain the rules or is that unnecesary?}

\paragraph{Clausification rules.}

In HOL, calculus rules can produce non-clausified results, necessitating repeated clausification. \sys{Leo-III} therefore lifts clausification to the calculus level. 
\EPrule{LiftEq}, \EPrule{CNFNeg}, \EPrule{CNFDisj}, \EPrule{CNFConj}, and \EPrule{CNFAll} lift HOL terms to the clause level, while \EPrule{CNFEx} performs Skolemization. In the implementation, the clausification procedure is complemented by a number of Boolean identities such as De Morgan’s laws.


\begin{minipage}{\textwidth}  
\raggedright
\begin{minipage}{0.9\textwidth}  
        \begin{minipage}{0.45\linewidth}
        \centering
              \begin{prooftree}
                \AxiomC{$\;C \lor \big[(l_{\tau}=r_{\tau}) \simeq \top\big]^{\alpha}\;$}
                \TightRuleLabel{\RuleDef{LiftEq}}
                \UnaryInfC{$\;C \lor \big[l_{\tau} \simeq r_{\tau}\big]^{\alpha}\;$}
              \end{prooftree}
            \end{minipage}
        \hfill
        \begin{minipage}{0.45\linewidth}
        \centering
              \begin{prooftree}
                \AxiomC{$\;C \lor \big[\neg s_{o}\big]^{\alpha}\;$}
                 \TightRuleLabel{\RuleDef{CNFNeg}}
                \UnaryInfC{$\;C \lor \big[s_{o}\big]^{\overline{\alpha}}\;$}
              \end{prooftree}
        \end{minipage}
\end{minipage}

\end{minipage}

\begin{minipage}{\textwidth}  
\raggedright

\begin{minipage}{0.9\textwidth}  
        \begin{minipage}{0.45\linewidth}
        \centering
            \begin{prooftree}
                \AxiomC{$\;C \lor \big[s_{o}\lor t_{o}\big]^{\pp}\;$}
                \TightRuleLabel{\RuleDef{CNFDisj}}
                \UnaryInfC{$\;C \lor \big[s_{o}\big]^{\pp}\lor \big[t_{o}\big]^{\pp}\;$}
              \end{prooftree}
            \end{minipage}
        \hfill
        \begin{minipage}{0.45\linewidth}
        \centering
             \begin{prooftree}
                \AxiomC{$\;C \lor \big[s_{o}\lor t_{o}\big]^{\nn}\;$}
                \TightRuleLabel{\RuleDef{CNFConj}}
                \UnaryInfC{$\;C \lor \big[s_{o}\big]^{\nn}\;$}
                \noLine
                \UnaryInfC{$\;C \lor \big[t_{o}\big]^{\nn}\;$}
              \end{prooftree}
        \end{minipage}
\end{minipage}

\begin{minipage}{0.9\textwidth}  
        \begin{minipage}{0.43\linewidth}
        \centering
           \begin{prooftree}
                \AxiomC{$\;C \lor \big[\forall X_{\tau}.\,s_{o}\big]^{\pp}\;$}
                \TightRightRuleLabel{\RuleDef{CNFAll}}{$Z_{\tau}$ is fresh}
                \UnaryInfC{$\;C \lor \big[s_{o}[Z_{\tau}/X]\big]^{\pp}\;$}
              \end{prooftree}
            \end{minipage}
        \hfill
        \begin{minipage}{0.47\linewidth}
        \centering
             \begin{prooftree}
                \AxiomC{$\;C \lor \big[\forall X_{\tau}.\,s_{o}\big]^{\nn}\;$}
                \TightRightRuleLabel{\RuleDef{CNFEx}}{$\mathit{sk}$ is a new\\Skolem term}
                \UnaryInfC{$\;C \lor \big[s_{o}[\,\mathit{sk} \; \overline{fv(C)} /X]\big]^{\nn}\;$}
              \end{prooftree}
        \end{minipage}
\end{minipage}

\end{minipage}


\paragraph{Extended calculus.}

\sys{Leo-III} complements this core calculus with a number of rules primarily motivated by the efficiency of the automation. Many of them delete or contract clauses or literals. A prominent example is \EPrule{Simp}, which applies 
Boolean identities to simplify terms: 

\noindent
\begin{minipage}{\textwidth}\centering
  \begin{prooftree}
    \AxiomC{$[l_1 \simeq r_1]^{\alpha_1} \lor \cdots \lor [l_n \simeq r_n]^{\alpha_n} \quad$}
    \RightLabel{\RuleDef{Simp}}
    \UnaryInfC{$[\text{simp}(l_1) \simeq \text{simp}(r_1)]^{\alpha_1} \lor \cdots \lor [\text{simp}(l_n) \simeq \text{simp}(r_n)]^{\alpha_n}$}
    \vspace*{1mm}
\end{prooftree}
\end{minipage}

\noindent
where $simp(t)$ exhaustively applies 17 identities, including the following:

\noindent
\begin{minipage}{\textwidth}\centering
\begin{tabular}{rl@{\hskip 1em}l@{\hskip 4em}rl@{\hskip 1em}l}
$s \land s$   & $\rightarrow s$ & ($\land_{idem}$)     &
$t \Rightarrow t$ & $\rightarrow \top$ & ($\Rightarrow_{refl}$) \\
$t \neq t$    & $\rightarrow \bot$ & ($\neq_{irrefl}$)    &
$s \lor \bot$ & $\rightarrow s$ & ($\lor\bot$) \\
\end{tabular}
\end{minipage}

The remaining rules of the extended calculus are given in 
\ifHALversion
  Appendix~\ref{ap:extendedCalculus}.
\else
  \cite{taprogge:hal-05735864}.
\fi
The full calculus includes 33 rules (3 primary inference rules, 5 unification rules, 4 extensionality rules, 6 clausification rules and 15 rules in the extended calculus) and is proved to be sound in \cite{LeoIIIthesis}.

\section{Encoding ATP Derivations}\label{sec:EncDeriv} 

ATPs for classical logics generally follow a refutation-based approach, where the negated conjecture is assumed. A proof then consists of a sequence of formulas derived from this negation (and any given axioms) via the calculus rules, ultimately leading to a contradiction ($\bot$). 
A general scheme for the proof-script encoding of the complete refutation proof of a formula \lpTerm{\emph{conjecture}} based on the individual steps encoded as given in Sec.~\ref{sec:encHOL}, is proposed in \cite{coltellacci2024reconstruction}:

\begin{center}
    \begin{lstlisting}
    symbol encoded_proof : π (*@\emph{conjecture}@*) ≔ 
    begin
        refine ¬¬ₑ (*@\emph{conjecture}@*) _;
        assume negated_conjecture;
        have step_1 :  Π x₁, ..., Π xₙ, π ...
            {assume x₁ ... xₙ; ...};
        ...
        have step_n : π ⊥ {...}
        refine step_n
    end;
    \end{lstlisting}
\end{center}

The script is initialized by \lpKeyword{begin} (line 2). 
The \lpKeyword{refine} tactic instantiates the proof-goal and is used 
to encode refutation proofs using double negation elimination (\lpTerm{$\neg\neg_e$}) (line 3), transforming the goal to \lpTerm{$\pi(\neg\neg$\emph{conjecture}$)$}, which reduces to  \lpTerm{$\pi$ ($\neg$ \emph{conjecture}$)\to$ $\pi$ $\bot$}. The \lpKeyword{assume} tactic allows reasoning under the hypothesis \lpTerm{negated\_conjecture : $\pi$($\neg$ \emph{conjecture})}, and leaves \lpTerm{$\pi$ $\bot$} as the goal. 
Each proof step is encoded as a lemma using the \lpKeyword{have} tactic (e.g., line 5), which does not add globally defined symbols but instead introduces local definitions within the context of the refutation proof. The steps are of shape \lpTerm{$\Pi \overrightarrow{x}.\,\varphi$} where \lpTerm{$\overrightarrow{x}$} denotes the universally quantified clause variables encoded using dependent types. The corresponding proof-scripts are given in curly braces.  At the beginning of the subproof, these variables are instantiated via \lpKeyword{assume}, which -- at the proof-term level -- introduces the corresponding $\lambda$-abstractions and thus binds the variables in the local context. The body of the subproof then uses representations of the calculus rules to provide a formal proof of the derivation within \sys{Lambdapi}, relying on the instantiated variables, the previous sub-steps, the encoded axioms, and \lpTerm{negated\_conjecture}.
The next subsection details the rule representations and how the \sys{Lambdapi} scripts invoke them.

\subsection{Encoding of Inference Rules}\label{subsection:encInfRules}

\begin{figure}[ht]
    \centering
    \begin{tikzpicture}[
      align=center,
      node distance=1cm and 1.5cm,
      arrow/.style={->, thick}
    ]
    
    \node[block] (structureQ) {Can the rule be represented by a native \sys{Lambdapi} operation?};

    \node[block, below=1.1cm of structureQ, xshift=3.5cm] (structure) {What part of the clause\\does the rule modify?};
    
    \node[block, below=1.1cm of structure, xshift=3.5cm] (substructure) {Relationship between \\ conclusion and premises};
    
    \node[subblockCyan, below=5cm of structureQ,  xshift=-2.8cm] (legend) {Encoding};
    \node[subblock, right=0.1cm of legend] (lpchoice) {-};
    \node[subblock, right=0.1cm of lpchoice] (function) {Function};
    \node[subblock, right=0.1cm of function] (consequence) {Function};
    \node[subblock, right=0.1cm of consequence] (equality) {Equality};
    
    \node[subblockCyan2, below=0.1cm of legend]{Application \\ to the clause};
    \node[subblock2, below=0.1cm of function]{Directly};
    \node[subblock2, below=0.1cm of consequence]{Via helpers};
    \node[subblock2, below=0.1cm of equality]{Via rewrite \\ tactic};

    \node[subblock2, below=0.1cm of lpchoice] (lpapply) {Via \sys{Lambdapi} \\ operation};
    
    \draw[arrow] (structure) -- node[midway, fill=white, inner sep=2pt, text centered] {Whole clause} (function);
    \draw[arrow] (structure) -- node[midway, fill=white, inner sep=2pt, text centered] {Substructure} (substructure);
    \draw[arrow] (substructure) -- node[midway, fill=white, inner sep=2pt, text centered, xshift=-10pt] {Consequence} (consequence);
    \draw[arrow] (substructure) --  node[midway, fill=white, inner sep=2pt, text centered, xshift=10pt] {Equivalence} (equality);

    \draw[arrow] (structureQ) -- node[midway, fill=white, inner sep=2pt] {no} (structure);
    \draw[arrow] (structureQ) -- node[midway, fill=white, inner sep=2pt] {yes} (lpchoice);

    \end{tikzpicture}
    
    \caption{Decision tree for choosing an encoding of an inference rule.}
    \label{fig:PrimInf}
\end{figure}
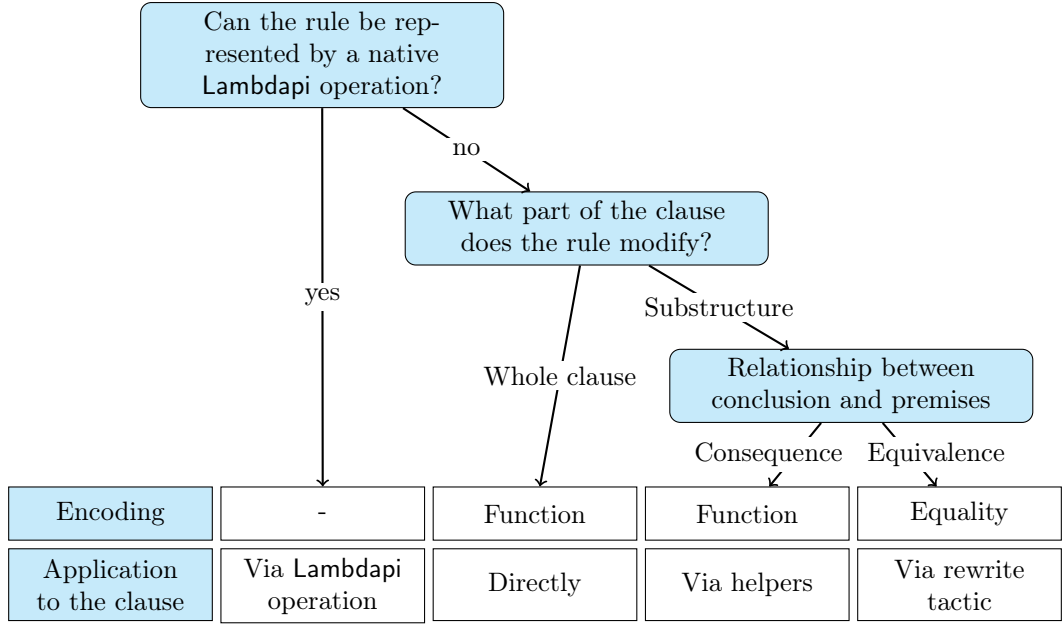

We categorize rules 
in terms of (a) the mechanisms they rely on (e.g., substitution or rewriting), (b) the clause (sub)structures they act on, and (c) whether -- in our encoding -- the conclusion is equivalent to the premises or merely a consequence. 
Fig.~\ref{fig:PrimInf} summarizes our encoding choices. We then outline and illustrate the approaches with examples.

Some \sys{Leo-III} rules correspond directly to \sys{Lambdapi} tactics. This covers, in particular, steps that are either essentially \emph{substitutions} (e.g., \RuleRef{Bind}), or \emph{rewriting}
.  In both cases, we do not need to introduce an explicit rule encoding, instead we directly use the corresponding \sys{Lambdapi} operations to derive a verifiable proof. In the first case, we instantiate the parent clause with the intended substitution. In the second case, we use the \lpKeyword{rewrite} tactic, which automatically produces proof-terms justifying the replacement of goal-subterms based on proven equalities. 
Other rules do benefit from an explicit encoding as \sys{Lambdapi} terms that (1) carry out the intended derivation and (2) are themselves proved valid in \sys{Lambdapi}. 
This encompasses rules operating on whole clauses, which can be encoded as functions and applied straightforwardly: 
using \lpKeyword{assume}, we first introduce instances of the universally quantified variables of the derived clause. These are then used to instantiate the rule and the clause variables from the previous step, 
allowing a direct mapping from the previous step to the one to be proved.
Application becomes more delicate for rules that target substructures (e.g.,\ rules acting on individual literals such as \RuleRef{PFE}). In these cases, we can still encode the rule as a function at the literal level, but have to combine it with helper lemmas that localize the transformation within the clause. 
When parent and child are logically equivalent (e.g., the Boolean identities applied by \RuleRef{Simp} and many clausification rules), the rules can be encoded as equalities and applied via the \lpKeyword{rewrite} tactic.

In the following, we demonstrate the use of our encoded rules but omit their proofs. The full encodings of all rules, 
together with examples of their application, are available in a GitHub repository~\cite{leo-lp-rep} and are reusable across systems. 

\subsubsection{Rules encoded as functions operating on substructures.}

The first example we consider is the rule \RuleRef{PFE}, which operates on individual literals and is encoded as the following function:

\begin{lstlisting}
symbol PFE [S] [T] (p q:τ(S ⤳ T)) x: π(p = q) → π(p x = q x);
\end{lstlisting}

The \lpTerm{PFE} symbol encodes the literal-level reasoning principle underlying \RuleRef{PFE}: from the equality of two functions, one can derive the equality of their applications to an argument. The full calculus rule, however, also extends the clause by a fresh universally quantified variable. This part is reflected not in the type of \lpTerm{PFE} itself, but in the generated proof-step encoding, which introduces the additional variable in the clause to be proved. 
For example, let \lpTerm{f} and \lpTerm{g} be terms of type \lpTerm{$\tau(S \leadsto T \leadsto U)$}:

\begin{lstlisting}
// assuming PFE_premise : Π x, π(f x = g x)
have PFE_example : Π x, Π y, π(f x y = g x y)
  {assume x y; refine PFE (f x) (g x) y (PFE_premise x)}
\end{lstlisting}

Here, \lpTerm{\lpKeyword{assume} x y} instantiates the universally quantified variables in the goal. The variable \lpTerm{x} corresponds to the variable already present in the premise, and \lpTerm{y} to the fresh variable introduced by \RuleRef{PFE}.
The term \lpTerm{PFE (f x) (g x) y} is then of type \lpTerm{$\pi$ (f x = g x) $\to$ $\pi$ (f x y =} \lpTerm{g x y)} and can thus be used to map \lpTerm{PFE\_premise x} to the desired proof. 
While we can use our encoding of \lpTerm{PFE} directly in this example, this is not necessarily the case for rules operating on substructures: If we consider, for instance, a premise \lpTerm{PFE\_premise': $\Pi$ x, $\pi$ (f x = g x $\lor$ l)} where the targeted literal is embedded in a disjunction with an additional literal \lpTerm{l}, we can no longer use \lpTerm{PFE} directly. 
Instead, we need additional \sys{Lambdapi} theorems as helpers. 
In the case of literal-level rules, this helper is the theorem \lpTerm{transform}, which is discussed below. 
The other EP core calculus rules encoded as functions over substructures also act at the literal level (\RuleRef{PBE}, \RuleRef{NBE}, \RuleRef{Fac}, and \RuleRef{Decomp}) and can therefore be applied with \lpTerm{transform}.

\subsubsection{Rules operating on clauses.}

Though not a rule of EP itself, \lpTerm{transform} is nevertheless an example of an inference operating on whole clauses and is accordingly encoded as a function. It can be represented as the following rule: 

\begin{prooftree}
    \AxiomC{$l_0 \lor l_1 \lor ... \lor l_i \lor ... \lor l_n$}
    \AxiomC{$l_i \Rightarrow l_i'$}
    \RightLabel{\RuleDef{Transform}}
    \BinaryInfC{$l_0 \lor l_1 \lor ... \lor l_i' \lor ... \lor l_n$}
    \singleLine
\end{prooftree}

Since clauses differ in length and the position of edited literals has to remain flexible, encoding \RuleRef{Transform} is more involved than the inferences already discussed. We therefore rely on the standard library encodings of natural numbers (of type \lpTerm{$\mathbb{N}$}) and lists (constructed via \lpTerm{$\mathbb{L}: \lpTerm{Set} \to \lpKeyword{TYPE}$}, indexed by encoded object-logic types). These allow for the representation of clauses as lists of literals (such as \lpTerm{c : $\mathbb{L}$ o}) and for the definition of functions like \lpTerm{nth} (to retrieve the n-th element of a list) and \lpTerm{set\_nth} (which replaces the n-th element of a list) defined via rewrite rules. These standard library encodings together with newly introduced ones, such as \lpTerm{disj} (which takes the disjunction of a list of terms of type \lpTerm{o}) allow us to state the following meta-theorem called \lpTerm{transform}:

\begin{lstlisting}
symbol transform [lᵢ' : Prop] (c: (*@$\mathbb{L}$@*) o) (i : (*@$\mathbb{N}$@*)):
 π((nth ⊥ c i) ⇒ lᵢ') → π(disj c) → π(disj(set_nth ⊥ c i lᵢ'));
\end{lstlisting}

Here, \lpTerm{$\bot$} is the default element used by \lpTerm{nth} and \lpTerm{set\_nth} for out-of-bounds indices. We can instantiate this theorem to prove \lpTerm{$\Pi$ x, $\Pi$ y, $\pi$ (f x y = g x y $\lor$ l)}: The first explicit argument is the list of literals of the clause we apply the rule to. With \lpTerm{${\colon\!\!\colon}$} denoting the list constructor, and \lpTerm{$\Box$} the empty list, \lpTerm{f x = g x $\lor$ l} is represented by \lpTerm{(f x = g x) ${\colon\!\!\colon}$ l ${\colon\!\!\colon}$ $\Box$}. The next two explicit arguments are the index of the transformed literal (\lpTerm{0}) 
and the rule allowing the transformation (\lpTerm{PFE (f x) (g x) y}). The application results in a function mapping the original clause \lpTerm{PFE\_premise'} to a version where \lpTerm{f x = g x} is replaced by \lpTerm{f x y = g x y}:

\begin{lstlisting}
// assuming PFE_premise' : Π x, π (f x = g x ∨ l)
have PFE_example' : Π x, Π y, π (f x y = g x y ∨ l)
 {assume x y; 
  refine transform ((f x = g x) ⸬ l ⸬ □) 0 (PFE (f x) (g x) y) 
      (PFE_premise' x)};
\end{lstlisting}


The remaining EP rules that append unification constraints (\RuleRef{FlexRigid}, \RuleRef{FlexFlex}, and \RuleRef{Prim}) are straightforward to encode, since their application can be verified using the standard library encodings \lpTerm{$\lor_{i1}$} and \lpTerm{$\lor_{i2}$} for the introduction rules of $\lor$. \RuleRef{INJ} operates on clauses with a fixed structure, and is directly encoded as a function, while those rules requiring structural flexibility (\RuleRef{DD} and \RuleRef{CNFConj}) rely on list-based encodings like the one demonstrated for \lpTerm{transform}.

\subsubsection{Rules encoded as equalities operating on substructures.}

Simplifications like $\land_{idem}$ target arbitrary substructures of clauses and therefore require greater flexibility. Neither of the methods introduced above can for instance be used to target \lpTerm{a $\land$ a} in \lpTerm{(a $\land$ a) = b}.  
Instead, we can use the \sys{Lambdapi} tactic \lpKeyword{rewrite} with rules encoded as equalities:

\begin{lstlisting}
symbol ∧_idem x: π ((x ∧ x) = x);
symbol ⇒_refl x: π ((x ⇒ x) = ⊤);
\end{lstlisting}
Rather than introducing new rewrite rules in the conversion, the \lpKeyword{rewrite} tactic constructs a proof-term justifying the replacement of any subterms matching the left-hand side of a proved equality with the corresponding right-hand side in the goal using the Leibniz principle. 
Optionally, a pattern can be supplied to specify the position of the term to be replaced.  
The effectiveness of this approach for handling simplifications has already been demonstrated in~\cite{coltellacci2024reconstruction}. 

\begin{lstlisting}
// assuming Simp_premise : π ((a ∧ a) = b)
have Simp_example : π (a = b) 
    {have SimpApp : π (((a ∧ a) = b) ⇒ (a = b)) 
        {rewrite ∧_idem; rewrite ⇒_refl; refine ⊤ᵢ}; 
    refine SimpApp Simp_premise};
\end{lstlisting}

Here, we first prove that the premise implies the simplified term as the substep \lpTerm{SimpApp}. The resulting goal \lpTerm{((a $\land$ a) = b) $\Rightarrow$ (a = b)} can be simplified using \lpTerm{$\land\_idem$} to obtain \lpTerm{(a = b) $\Rightarrow$ (a = b)}. 
Simplification via \lpTerm{$\Rightarrow$\_refl} reduces the goal to \lpTerm{$\top$}. In \sys{Lambdapi}, \lpTerm{$\top$} is proved by \lpTerm{$\top_i$}, which is therefore used to close the subproof. Finally, we apply \lpTerm{SimpApp} to 
\lpTerm{Simp\_premise} to derive the desired conclusion.
In our encoding, over 40 equalities are used to encode the simplifications applied by \sys{Leo-III}, the encodings of which have been proved and added to the \sys{Lambdapi} standard library. 
In the example above, we show the corresponding rewrite steps explicitly. In the implementation, however, we use a user-defined \sys{Lambdapi} tactic that applies the available simplification equalities exhaustively to the current goal and then closes it with \lpTerm{$\Rightarrow\_refl$} and \lpTerm{$\top_i$} once it has been reduced to a reflexive implication.

\paragraph{Skolemization.}
While Skolemization generally preserves only satisfiability, in HOL with Hilbert choice the Skolem symbols introduced during clausification can be defined in terms of the choice operator. In our encoding, each Skolem symbol is defined as a notation for an underlying $\varepsilon$-term, and the corresponding transformation is justified by the following rule:

\begin{lstlisting}
symbol CNFEx (a: Set) (p: τ a → Prop) :  
  π (¬ (∀ p) = ¬ (p ((*@$\varepsilon$@*) (λ x, ¬ (p x)))))
\end{lstlisting}

During clausification, \sys{Leo-III} then generates a definition for each introduced Skolem symbol, e.g., 
\lpTerm{sk1 $\coloneqq$ $\varepsilon$ ($\lambda$ x, $\neg$ (p x))}, which can be applied to unfold the named Skolem terms in the goal using the \lpKeyword{simplify} tactic:


\begin{lstlisting}
// assuming sk1 ≔ (*@$\color{red}{\varepsilon}$@*) (λ x, ¬ (p x))
// and CNFEx_premise: π (c ∨ ¬(∀ p))
have CNFEx_example: π (c ∨ ¬ (p sk1))
    {have Skolemization : π (c ∨ ¬(∀ p) ⇒ c ∨ ¬ (p sk1))
        {rewrite CNFEx; simplify sk1; 
         rewrite ⇒_refl; refine ⊤ᵢ};
    refine Skolemization CNFEx_premise }
\end{lstlisting}

The other EP rules verified through such equality-based encodings are \RuleRef{NFE}, \RuleRef{CNFAll}, and the Boolean identities used throughout clausification.

\subsubsection{Rules corresponding to \sys{Lambdapi} operations.}

\RuleRef{Bind} resolves unification constraints like $[s = t]^{\nn}$ by applying a substitution $\sigma$ with $\sigma(s)=\sigma(t)$ to the entire clause and dropping the constraint.
In our encoding, this amounts to (i) instantiating the universally quantified variables of the parent clause in accordance with $\sigma$ and (ii) removing the now-trivially-false constraint. The latter is achieved through the sequential application of two simplifications also covered by \RuleRef{Simp} ($\neq_{irrefl}$ and $\lor\bot$).
Once again, these rules are encoded as equalities: $\neq_{irrefl}$ is encoded as \lpTerm{$\neg=_{irrefl}$} and  two \sys{Lambdapi} theorems are given for $\lor\bot$:  \lpTerm{$\bot\lor$} and \lpTerm{$\lor\bot$}, for $\bot\lor x = x$ and $x\lor\bot = x$ respectively. 
In this fashion, we can 
verify applications of \RuleRef{Bind}:

\begin{lstlisting}
// assuming Bind_premise : Π x, Π y, π (f x y ∨ ¬(x = a))
have Bind_example : Π y, π (f a y)
 {assume y;
 have Substitution : π (f a y ∨ ¬(a = a))
    {refine Bind_premise a y};
  have RemoveConstraint : π ((f a y ∨ ¬(a = a)) ⇒ (f a y))  
    {rewrite ¬=_irrefl; rewrite ∨⊥; rewrite ⇒_refl; refine ⊤ᵢ};
  refine RemoveConstraint Substitution}
\end{lstlisting}

Here, \lpTerm{Substitution} performs the instantiation, and \lpTerm{RemoveConstraint} applies the remaining rewrites to discharge the trivial literal.

\RuleRef{Triv} removes trivially false literals from clauses and is simply encoded by literal deletion, analogously to \RuleRef{Bind}. \RuleRef{Para}, on the other hand, represents conditional rewriting and can be encoded using the \sys{Lambdapi} \lpKeyword{rewrite} tactic.

\subsection{Flexible Encoding of Proof Steps}

In order to apply encoded inference rules in the verification of actual derivation steps, one more factor has to be considered: systems can introduce slight modifications to the derived formulas resulting from the implementation of the prover itself. This phenomenon has also been observed in the verification of other systems \cite{coltellacci2024reconstruction,komel2025case}. We discuss such \emph{implicit transformations} and an encoding approach that integrates them into the verification process in the following.

\paragraph{Implicit transformations.}

The possible implicit transformations can vary between provers and calculus rules and need to be identified via the analysis of the implementation of the prover. We will demonstrate the nature and handling of such transformations using the example of \RuleRef{PFE}. 
In the application of the rule, \sys{Leo-III} first separates the literals to which the rule can be applied (i.e., equalities between functions) from the rest. It then applies the same fresh variable to both sides of the equalities individually, and, in forming new literals, rearranges the sides in accordance with a term ordering. The resulting clause is formed by the unchanged literals followed by the transformed ones. 
It is thus possible that an application of \RuleRef{PFE} changes both the sides within equality literals and the order of the literals within the clause. In fact, both of these transformations would occur when applying \RuleRef{PFE} to the clause we encoded as \lpTerm{PFE\_premise'}. To account for them in our encoding, we hence need to derive a proof of \lpTerm{$\Pi$ x, $\Pi$ y, $\pi$ (l $\lor$ g x y = f x y)} rather than \lpTerm{$\Pi$ x, $\Pi$ y, $\pi$ (f x y = g x y $\lor$ l)}.
Generally, implicit transformations introduced by a prover can themselves be seen as inference rules applied in addition to the calculus rules. We can thus treat implicit transformations like calculus rules and derive an encoding for them using the schema given in Fig.~\ref{fig:PrimInf}. For the reordering of terms within equalities, this results in the encoding of the rule as an equality called \lpTerm{eqSym}, and, for permutations, in a function called \lpTerm{permute}:

\begin{lstlisting}
symbol eqSym [T] (x y : τ T) : π((x = y) = (y = x));

symbol permute (σ : (*@$\mathbb{L}$@*) nat) (c: (*@$\mathbb{L}$@*) o) : 
  π(preserves_contents σ c) → π(disj c) → π(disj(literals c σ));
\end{lstlisting}

\noindent
\lpTerm{permute} can be instantiated with a list of literals of the original clause \lpTerm{c}, and a list of natural numbers \lpTerm{$\sigma$} representing the desired permutation. 
The function \lpTerm{preserves\_contents} requires that all literals present in \lpTerm{c} are still contained in the clause after the proposed permutation \lpTerm{$\sigma$}. This subgoal is solved by mere computation following the approach of reflexive tactics \cite{DBLP:conf/tacs/Boutin97}, and reduces to \lpTerm{$\pi \top$} if the condition is met.  It can thus be instantiated with the standard library introduction rule \lpTerm{$\top_i$}. The resulting function maps a proof of the disjunction of \lpTerm{c} to its permuted version, which is computed via rewrite rules by \lpTerm{literals c $\sigma$}. 




\paragraph{A flexible encoding schema.}


Note that some implicit transformations must be carried out before the verification of the actual calculus rule (e.g., when a particular order of literals is required), while others can only be applied afterward.
Whether or not a given implicit transformation that can occur for a calculus rule application is necessary in a concrete generated proof must be decided on a case-by-case basis. The following flexible encoding schema can thus be applied to encode rule applications, including any potential additional transformations:

\vspace*{-2mm}
\begin{enumerate}[noitemsep, label=\arabic*.]
    \item Assume the quantified variables.
    \item Apply any implicit transformations necessary prior to the calculus rule.
    \item Apply the calculus rule.
    \item Apply the remaining implicit transformations.
    \item Instantiate the goal.
\end{enumerate}
\vspace*{-1mm}

Following this strategy for \RuleRef{PFE}, we first \lpKeyword{assume} any quantified variables (Step 1) and then apply \lpTerm{PFE} directly (Step 3). The implicit transformations (reordering within equality literals and permutation) are detected during proof-script construction by \sys{Leo-III} and, if necessary, handled after the calculus rule application (Step 4). 
Finally, the goal is instantiated (Step 5).
This way, we can at last prove \lpTerm{$\Pi$ x, $\Pi$ y, $\pi$ (l $\lor$ g x y = f x y)}:

\begin{center}
\begin{lstlisting}
// assuming PFE_premise' : Π x, π (f x = g x ∨ l)
have PFE_example'' : Π x, Π y, π (l ∨ g x y = f x y)
 {assume x y;  // Step 1
  have PFE_app: π (f x y = g x y ∨ l)  // Step 3
    {refine transform ((f x = g x) ⸬ l ⸬ □) 0 (PFE (f x) (g x) y) (PFE_premise' x)};
  rewrite eqSym (g x y) (f x y);  // Step 4
  have Permutation: π (l ∨ (f x y = g x y))
    {refine permute (1 ⸬ 0 ⸬ □) ((f x y = g x y) ⸬ l ⸬ □) ⊤ᵢ PFE_app};
  refine Permutation}   // Step 5
\end{lstlisting}
\end{center}

\section{Evaluation}\label{sec:Eval}

\def\totalSolvableProblems{1803}
\def\totalSolvableProblemsLP{1784}
\def\totalSteps{35578}
\def\totalEncodedSteps{28512}
\def\coverage{80\%}
\def\avgRunTimeLeo{2.00\,s}
\def\avgRunTimeLeoLP{2.11\,s}
\def\avgTSTPsize{13074\,B}
\def\avgLPproofSize{19206\,B}
\def\avgLPpcheckTime{0.76\,s}

The implementation of the verification approach presented in this work involves three key steps: (i) deriving a theoretical encoding of each of the calculus rules of the prover, (ii) identifying 
implicit transformations based on the prover's implementation, and integrating them in a flexible encoding
, and (iii) implementing the automated 
encoding of proof steps following the resulting schema. 
Furthermore, the original code may need to be modified to record additional details, such as term positions.
Task (i) is complete: Table~\ref{tab:categories} classifies all rules of the EP calculus with respect to the encoding mechanisms discussed in Sec.~\ref{subsection:encInfRules} and indicates whether automated generation of \sys{Lambdapi} proofs is currently implemented. A corresponding table for the rules extending the EP calculus in \sys{Leo-III} is given in
\ifHALversion
  Appendix~\ref{ap:extendedCalculus}.
\else
  \cite{taprogge:hal-05735864}.
\fi
Note that the rules \RuleRef{LiftEq}, \RuleRef{CNFNeg}, and \RuleRef{CNFDisj} perform transformations abstracted away by the \sys{Lambdapi} encoding. We therefore only need to encode the implicit transformations they introduce. 
All encodings, illustrated by examples, are available online \cite{leo-lp-rep}.

\newcommand{\FunAddUniCon}{Function operating on clauses (\lpTerm{$\lor_1$} and \lpTerm{$\lor_2$})}
\newcommand{\FunOnClause}{Function operating on clause}
\newcommand{\corrRW}{Corresponding \sys{Lambdapi} operation (\lpKeyword{rewrite})}
\newcommand{\corrInst}{Corresponding \sys{Lambdapi} operation (instantiation)}
\newcommand{\FunOnSubst}{Function operating on substructures}
\newcommand{\RwEq}{Encoded as equation, applied via \lpKeyword{rewrite}}
\newcommand{\onlySideEff}{Only side-effects need verification}
\newcommand{\omitRule}{\textit{needs no encoding}}

\begin{table*}[!htbp]
\centering
\caption{Encoding categories for the EP core calculus.}
\label{tab:categories}
\begin{tabular}{@{}>{\centering\arraybackslash}m{1cm} | l l c@{}}
\toprule
\textbf{Cat.} & \textbf{Rule} & \textbf{Encoding category} & \textbf{Implemented} \\
\midrule
\tablecat{Primary\\rules}{3} & \RuleRef{Prim}      & \FunAddUniCon & No \\
 & \RuleRef{Para}      & \corrRW & Yes \\
 & \RuleRef{Fac}       & \FunOnSubst & Yes \\
\midrule
\tablecat{Unification\\rules}{5} & \RuleRef{Triv}      & \RwEq & Yes \\
 & \RuleRef{Bind}      & \corrInst & Yes \\
 & \RuleRef{Decomp}    & \FunOnSubst & Yes \\
 & \RuleRef{FlexRigid} & \FunAddUniCon & No \\
 & \RuleRef{FlexFlex}  & \FunAddUniCon & No \\
\midrule
\tablecat{Extens.\\rules}{4}  & \RuleRef{PFE} & \FunOnSubst & Yes \\
 & \RuleRef{NFE} & \RwEq & Yes \\
 & \RuleRef{PBE} & \FunOnSubst & Yes \\
 & \RuleRef{NBE} & \FunOnSubst & No \\
\midrule
\tablecat{Clausification\\rules}{6} & \RuleRef{CNFEx}    & \RwEq & Yes \\
 & \RuleRef{LiftEq}   & \onlySideEff & Yes \\
 & \RuleRef{CNFNeg}   & \onlySideEff & Yes \\
 & \RuleRef{CNFDisj}  & \onlySideEff & Yes \\
 & \RuleRef{CNFConj}  & \FunOnClause & Yes \\
 & \RuleRef{CNFAll}   & \RwEq & No \\
\bottomrule
\end{tabular}
\end{table*}

Tasks (ii) and (iii) are addressed by our implementation\footnote{Available on \url{https://github.com/melanie-taprogge/Leo-III}.}, an open-source extension of \sys{Leo-III} that currently focuses on monomorphic input problems. It outputs \sys{Lambdapi} scripts proving supported steps, and uses the tactic \lpKeyword{admit} to assume the remaining steps as axioms. 
The prototype covers instances of 13 of the 18 core calculus rules across all categories, omitting some edge cases. 
For the extended calculus, 3 of 11 relevant rules are currently covered (see~
\ifHALversion
  Appendix~\ref{ap:extendedCalculus}
\else
  \cite{taprogge:hal-05735864}
\fi
for details). 
The implementation of the automated generation of proof-scripts is highly labor-intensive. It requires precise analysis of the original code and, in many cases, changes to the original implementation of \sys{Leo-III} in order to retrieve the level of detail needed for reconstruction in \sys{Lambdapi}. 
The implementation of automated \sys{Lambdapi} output must then reliably identify and justify not only the exact variant of each applied calculus rule, but also all required implicit transformations. 
Our current partial implementation adds approximately 5500 lines of code for proof-script construction to the \sys{Leo-III} code base, corresponding to about 14\% of the roughly 38500 lines of code in the regular code base excluding \sys{Lambdapi} proof output.  
The encoding furthermore builds on a large number of theorems and lemmas proved in \sys{Lambdapi} for representing calculus rules and implicit transformations. More than 1500 lines of \sys{Lambdapi} code, comprising over 80 reusable theorems, were added to the \sys{Lambdapi} standard library, while our dedicated library of \sys{Leo-III}-specific encodings comprises over 600 lines of code and a further 35 theorems. 
While the automated encoding of proof steps has not been completed for the full calculus yet, we have successfully constructed \sys{Lambdapi} scripts for hand-crafted examples for all remaining rules and are confident that we will be able to verify them using our approach and encodings. 
The main technical difficulty in the implementation is (a) covering edge cases -- such as rule-specific implicit transformations -- and (b) handling procedures that apply many rules blindly in a normalization phase, as is the case for clausification. While \sys{Lambdapi}’s user-defined tactics and handling of implicit arguments let us avoid logging every detail during proof search and, in some cases, allow blind rule applications mirroring processes implemented in \sys{Leo-III}, the generation of verifiable proofs becomes delicate when heterogeneous rules are bundled. For instance, \sys{Leo-III} combines all clausification operations into a single step, so even one unsupported occurrence of \RuleRef{CNFAll} can prevent verification of the entire step, even when all other needed clausification operations are encodable.

To evaluate our current implementation, we used the 3951 monomorphic HOL problems from the TPTP-v9.2.1 library \cite{TPTPplatform} as a benchmark, of which \sys{Leo-III} solved \totalSolvableProblems{} within a 60-second time limit in its default configuration. Under the same time limit, the number of problems for which the extended version produced Lambdapi proofs was slightly lower (\totalSolvableProblemsLP{}).  
This reduction is due to the additional overhead of proof formalization, which is to be expected and also observed in related work~\cite{komel2025case}, resulting in slightly fewer problems solved under the same time limit. 
In total, the generated proofs included \totalSteps{} steps, of which \totalEncodedSteps{} were reconstructed as checkable \sys{Lambdapi} proof scripts, yielding an automatic reconstruction rate of approximately \coverage{}. All generated proof scripts were accepted by \sys{Lambdapi}. 
The remaining 20\% are, for the most part, 
steps originating from bundled calculus rule applications, or unencoded applications of rules from the extended calculus, both of which will be addressed in future work. 
However, a significant part of the admitted steps is, in principle, already covered by our equality-based encodings but requires the application of the \lpKeyword{rewrite} tactic under binders. This is not currently supported by \sys{Lambdapi}, a restriction that will likewise be addressed in future work. 
\melaniem{only 78 are fully verifiable, a total of 309 are verifiable up to clausification}
On average, our Leo‑III version with \sys{Lambdapi} output required \avgRunTimeLeoLP{}, compared to \avgRunTimeLeo{} for the standard version, while proof checking took an average of \avgLPpcheckTime{}. The \sys{TSTP} proofs output by \sys{Leo-III} 
had an average size of \avgTSTPsize{}, whereas the corresponding \texttt{.lp} files 
averaged \avgLPproofSize{}. The results are summarized in Table~\ref{tab:performanceMetrics}.

This implementation effort also revealed several bugs in \sys{Leo-III}, illustrating the practical benefits of proof verification. Some proof certificates were incomplete, as they omitted type declarations or TPTP definitions required to fully reconstruct individual proof steps. We also found cases where defined TPTP connectives were unfolded unintentionally, producing formulas in which relevant connective structure was no longer visible to parts of the calculus. Finally, the formalization exposed an issue in one of the unification rules that could lead to invalid unifications and therefore to potentially unsound proof steps.

\begin{table}[ht]
\centering
\caption{Performance Metrics Overview}
\label{tab:performanceMetrics}

\noindent\makebox[\linewidth][c]{%
\begin{minipage}[t]{0.44\linewidth}%
\centering%
\textbf{Coverage \& Proof Statistics}\strut\\[4pt]%
{\setlength{\tabcolsep}{4pt}\renewcommand{\arraystretch}{1.1}%
\begin{tabular}{@{}lr@{}}%
\toprule%
\textbf{Metric} & \textbf{Value} \\%
\midrule%
Proofs (\sys{Leo-III}+TSTP-out.) & \totalSolvableProblems{} \\%
Proofs (\sys{Leo-III}+LP-out.) & \totalSolvableProblemsLP{} \\%
Total proof steps & \totalSteps{} \\%
Encoded steps & \totalEncodedSteps{} \\%
Encoding coverage & \coverage{} \\%
\bottomrule%
\end{tabular}}%
\end{minipage}%
\hspace{.02\linewidth}%
\begin{minipage}[t]{0.49\linewidth}%
\centering%
\textbf{Runtime \& Proof Sizes}\strut\\[4pt]%
{\setlength{\tabcolsep}{4pt}\renewcommand{\arraystretch}{1.1}%
\begin{tabular}{@{}lr@{}}%
\toprule%
\textbf{Metric} & \textbf{Value} \\%
\midrule%
Avg.\ runtime (\sys{Leo-III}+TSTP-out.) & \avgRunTimeLeo{} \\%
Avg.\ runtime (\sys{Leo-III}+LP-out.) & \avgRunTimeLeoLP{} \\%
Avg.\ \sys{Lambdapi} checking time & \avgLPpcheckTime{} \\%
Avg.\ \sys{TSTP} proof size & \avgTSTPsize{} \\%
Avg.\ \sys{Lambdapi} proof size & \avgLPproofSize{} \\%
\bottomrule%
\end{tabular}}%
\end{minipage}%
}
The benchmarks were run on an Ubuntu 24.04.3 LTS machine with 16 vCPUs\\and 64 GB RAM on an AMD Ryzen 9 9950X host CPU.
\end{table}

\vspace{-2em}





\section{Conclusion}\label{sec:Concl}

In this work, we present a systematic approach 
for the encoding of automated reasoning in \sys{Lambdapi} and apply it to the HOL ATP \sys{Leo-III}. 
The resulting implementation already provides independently checkable reconstruction for a large share of proof steps occurring in real \sys{Leo-III} proofs. 
Our theoretical encoding of all calculus rules demonstrates that 
our encoding identifies solutions to the main conceptual challenges.
Notably, this includes handling clausification steps, which are often omitted from verification efforts (e.g. \cite{komel2025case,iProver}). The challenges commonly encountered in their verification are (i) Skolemization, which generally preserves only equisatisfiability, and (ii) the fact that ATPs typically do not track each applied clausification rule. 
In our encoding, we overcome these hurdles by relying on (i) the use of $\varepsilon$-terms and (ii) the \lpKeyword{rewrite} tactic together with \sys{Lambdapi}’s capability to infer instantiation of rules and the term positions to which rules must be applied. 
The general approach presented here is directly applicable to other ATPs, and the encoded rules are reusable by other HOL systems implementing calculi that share rules with EP. The work that remains to be done for each individual system is the encoding of system-specific rules following our encoding scheme, the detection of implicit transformations caused by the implementation of the prover, and the implementation of the automated generation of \sys{Lambdapi} scripts. 
Beyond HOL systems, the strategies we introduce address practical challenges also observed in related projects -- such as literal permutations or the reordering of equality sides \cite{komel2025case,coltellacci2024reconstruction} and literals in clauses \cite{coltellacci2024reconstruction} -- which necessitate explicit justification during verification. These works handle such transformations by proof elaboration via external tools and by the construction of dedicated explicit proof terms, respectively. In contrast, our framework addresses these transformations via reusable \sys{Lambdapi} theorems and tactics that can be applied uniformly across systems in a straightforward manner.

\melaniem{is this a sufficient justification for "why should we publish now and not wait until coverage is at 100\%?"}
\fredm{I think so}
\alexm{One could maybe add something in the lines of ,,the remaining stuff is just engineering effort but essentially solved''?}
\fredm{I prefer not to say that for not depreciating the interest of the next paper, all the more so since we don't know until we are actually doing it.}

In future work, we will incorporate encodings for polymorphism and complete the rule coverage. 
This will not only make \sys{Leo-III} the first HOL ATP system capable of outputting \sys{Lambdapi}, but -- to our knowledge -- also the first mature HOL ATP capable of directly translating its proofs to an independently verifiable formalism. 
The integration of \sys{Leo-III} in the \sys{Dedukti} framework will also enable the verification of proofs in a range of non-classical logics for the first time, which \sys{Leo-III} handles through a built-in shallow embedding of various non-classical logics into HOL. 
Furthermore, an end-to-end translation pipeline from \sys{Leo-III} proofs via \sys{Lambdapi} to \sys{Rocq} is under development. 
Finally, we plan an integration with the \sys{GDV-LP} tool \cite{DBLP:conf/flairs/SutcliffeBB25}, a version of \sys{GDV} relying on \sys{Lambdapi}-proof-producing ATPs.
\sys{Leo-III} will be able to lift the limitation of the current tool to FOL.
A \sys{Lambdapi}-producing version of \sys{Nörgler} relying on a \sys{Leo-III} integration is also planned. 


\bibliographystyle{plain}
\bibliography{ref}

\ifHALversion
\appendix

\newpage

\section{Translation and Meta-Properties} \label{ap:HOLenc} 

Here, we present the formal translation of the HOL types, constants, terms, and problems into \sys{Lambdapi}. We also recall the arguments for the decidability of type-checking, and the correctness of the encoding.

\subsection{Encoding of HOL}

All declarations and rewrite rules below are taken from the \sys{Lambdapi} standard library\footnote{\url{https://github.com/Deducteam/lambdapi-stdlib}}.

\begin{definition}[Signature $\Sigma$]
$\Sigma$ contains the declarations of all encoded HOL logical symbols, which are given below.
\end{definition}

\begin{center}
\begin{lstlisting}
// Encoding of HOL types
constant symbol Set : TYPE;
constant symbol ι : Set;
constant symbol o : Set;
constant symbol ⤳ : Set → Set → Set; 
injective symbol τ : Set → TYPE;

// Encoding of HOL connectives
constant symbol Prop : TYPE;
constant symbol ⊤ : Prop; 
constant symbol ⊥ : Prop; 
constant symbol ⇒ : Prop → Prop → Prop; 
constant symbol ∨ : Prop → Prop → Prop; 
constant symbol ∧ : Prop → Prop → Prop; 
symbol ¬ p ≔ p ⇒ ⊥;
symbol ⇔ p q ≔ (p ⇒ q) ∧ (q ⇒ p);
constant symbol ∀ [a] : (τ a → Prop) → Prop; 
constant symbol ∃ [a] : (τ a → Prop) → Prop; 
constant symbol = [a] : τ a → τ a → Prop;
injective symbol π : Prop → TYPE; 
\end{lstlisting}
\end{center}

\begin{definition}[Rewrite rules $\mathcal{R}$]
The set of rewrite rules associated with the signature  is denoted by $\mathcal{R}$ and can be found below.
\end{definition}

\begin{center}
\begin{lstlisting}
rule τ ($x ⤳ $y) ↪ τ $x → τ $y;
rule τ o ↪ Prop;
rule π ($p ⇒ $q) ↪ π $p → π $q;
rule π (∀(λ x, $f.[x])) ↪ Π x, π $f.[x];
\end{lstlisting}
\end{center}


\begin{definition}[Natural Deduction rules $\mathcal{ND}$]
$\mathcal{ND}$ comprises the introduction and elimination rules for the logical symbols of $\Sigma$. No rules for $\forall$ and $\neg$ are included, as they are derivable from the rewrite rules in $\mathcal{R}$. The rules for the remaining symbols are given below.
\end{definition}

\begin{center}
\begin{lstlisting}
constant symbol ⊤ᵢ : π ⊤;

constant symbol ⊥ₑ [p] : π ⊥ → π p;

constant symbol ∧ᵢ [p q] : π p → π q → π (p ∧ q);
symbol ∧ₑ₁ [p q] : π (p ∧ q) → π p;
symbol ∧ₑ₂ [p q] : π (p ∧ q) → π q;

constant symbol ∨ᵢ₁ [p q] : π p → π (p ∨ q);
constant symbol ∨ᵢ₂ [p q] : π q → π (p ∨ q);
symbol ∨ₑ [p q r] : π (p ∨ q) → (π p → π r) → (π q → π r) → π r;

constant symbol ∃ᵢ [a p] (x:τ a) : π (p x) → π (∃(λ x, p x)); 
symbol ∃ₑ [a p]: π(∃(λ x, p x))→ Π [q], (Π x:τ a, π(p x) → π q) → π q;
\end{lstlisting}
\end{center}

\begin{definition}[Axiomatization $\mathcal{A}$]
The metalogical assumptions underlying our encoding are the non-emptiness of object-level types (\lpTerm{el}), reflexivity and inductive elimination of equality (\lpTerm{eq\_refl} and \lpTerm{ind\_eq}), excluded middle (\lpTerm{em}), functional and propositional extensionality (\lpTerm{funExt} and \lpTerm{propExt}) and the introduction of Hilbert Choice (\lpTerm{$\varepsilon_i$}). Their encodings (given below) form $\mathcal{A}$.
\end{definition}

\begin{center}
\begin{lstlisting}
symbol el a : τ a; // types are not empty

constant symbol eq_refl [a] (x:τ a) : π (x = x);
symbol ind_eq [a] [x y:τ a] : π (x = y) → Π p, π (p y) → π (p x);

symbol em p : π (p ∨ ¬ p); // excluded middle

symbol funExt [a b] (f g: τ (a ⤳ b)): (Π x, π(f x = g x))→ π(f = g);
symbol propExt x y : (π x → π y) → (π y → π x) → π (x = y);

symbol (*@$\varepsilon$@*) [a:Set] : τ ((a ⤳ o) ⤳ a); // choice
symbol (*@$\varepsilon$@*)ᵢ [a:Set] (p:τ a → Prop) : π (∃ p) → π (p ((*@$\varepsilon$@*) p));
\end{lstlisting}
\end{center}

\begin{definition}[System $\mathcal{HOL}$]
The tuple $(\Sigma,\mathcal{R},\mathcal{ND},\mathcal{A})$ is defined as \textbf{System $\mathcal{HOL}$}.
\end{definition}

\subsection{Translation}

\begin{definition}[Embedding of Types and Constants]\label{def:translateLang}
The \textbf{deep embedding} of types $T$, denoted by $|.|$, is given recursively as:
\[
|T| = \text{\lpTerm{$T$}}, \quad
|T_1 \rightarrow T_2| = \text{\lpTerm{$|T_1| \leadsto |T_2|$}}.
\]
The corresponding \textbf{shallow embedding}, denoted by $\|. \|$, is defined as $\|T\| = \text{\lpTerm{$\tau$ $|T|$}}$. 
For a HOL language $\mathcal{L}$, each \textbf{type constant} $T \in \mathcal{L}$ is declared as \text{\lpTerm{$T$ : Set}}, 
and each \textbf{term constant} $c : T \in \mathcal{L}$ is declared as \text{\lpTerm{$c$ : $\|T\|$}}.
\end{definition}

\begin{definition}[Term Embedding] \label{def:translateTerms}
The embedding of object-level terms, denoted by $|.|$, is defined inductively as follows:
\allowdisplaybreaks

\begin{align*}
    |c| = \text{\lpTerm{$c$}}, \quad
    |x| & = \text{\lpTerm{x}}, \quad
    |t\,s| = \text{\lpTerm{$|t|\,|s|$}}, \quad
    |\lambda x_T, t| = \text{\lpTerm{$\lambda$x:$\|T\|$,$|t|$}}.
\end{align*}

\centering
\footnotesize{Here, $c$ denotes a constant and $x$ a variable.}

\end{definition}

\begin{definition}[Problem Embedding] \label{def:translateProblems}
A proposition $\gamma$ is encoded as $\|\gamma\| = \text{\lpTerm{$\pi |\gamma|$}}$. Let $\Delta = \{\gamma_1,....\gamma_n\}$ be the set of axioms of a HOL reasoning problem, $(x_{i,1} : T_{i,1}),...,(x_{i,m} : T_{i,m})$ the universally quantified variables of $\gamma_i$ and $p_i$ a fresh constant for every $i$. $\Delta$ is then embedded as \lpTerm{p$_1$:$\Pi x_{1,1} : \|T_{1,1}\|$,$ \,...\, $,$\Pi x_{1,m} : $} \lpTerm{$\|T_{1,m}\|$,$ \, \|\gamma_1\|$, ...,p$_n$:$\Pi x_{n,1} : \|T_{n,1}\| $,$ \,...\, $,$ \Pi x_{n,m} : \|T_{n,m}\| $,$ \, \|\gamma_n\|$}.
\end{definition}

\subsection{Decidability of Type-Checking}\label{subSec:DecTyping} 

It is shown in \cite{assaf2015framework} that type-checking in $\lambda\Pi$-calculus modulo theory for a given theory is decidable if (1) the left-hand sides of all rewrite-rules are restricted to Miller's patterns \cite{DBLP:journals/logcom/Miller91}, which is enforced in the Dedukti framework, and (2) the $\to_{\beta\mathcal{R}}$ relation satisfies three properties: \emph{confluence}, \emph{subject reduction} and \emph{termination}.
Confluence guarantees that a term can have at most one irreducible form. Specifically, given $\mathcal{R}$ (the set of rewrite rules of the system $\mathcal{HOL}$) and $\rightarrow_{\beta \mathcal{R}}^*$ (the reflexive-transitive closure of $\rightarrow_{\beta \mathcal{R}}$), the relation $\rightarrow_{\beta \mathcal{R}}$ is confluent if, for all terms $t$, $s_1$, and $s_2$, the condition $(t \rightarrow_{\beta \mathcal{R}}^* s_1 \land t \rightarrow_{\beta \mathcal{R}}^* s_2) \; \Rightarrow \; \exists t'. (s_1 \rightarrow_{\beta \mathcal{R}}^* t' \land s_2 \rightarrow_{\beta \mathcal{R}}^* t')$ holds. 
It is proved for our encoding in \cite{systemU} based on the absence of \emph{critical pairs} among the rewrite rules, combined with their left-linearity \cite{klop1993combinatory}.

From confluence, together with the requirement that the left- and right-hand sides of all rewrite rules are typable and share the same type, a condition that holds in our encoding, subject reduction follows. This states that if a term $t$ reduces to $t'$ under $\to_{\beta\mathcal{R}}$, then $t'$ retains the same type as $t$.

Lastly, termination ensures that any reduction process must terminate after a finite number of steps. It has been demonstrated in \cite{Dowek17termination} for minimal HOL and extended to the constructive and classical HOL fragments of the encoding we follow in \cite{DBLP:conf/types/Grienenberger22}. 
Our encoding incorporates equality, its defining properties, and HOL's foundational axioms through additional declarations, without introducing new rewrite rules. Consequently, the proof remains valid for our encoding.

Together, confluence, termination and type preservation guarantee the existence of a normal form for any term, enabling the comparison of terms with respect to $\rightarrow_{\beta \mathcal{R}}$ and thereby making type-checking decidable.

\subsection{Soundness and Conservativity of the Encoding}\label{subSec:decidability} 



The correctness of the encoding ensures a precise correspondence between terms, statements, and proofs in the original system and their encoded counterparts. Establishing correctness requires both soundness and conservativity. Soundness guarantees that every ND proof in HOL can be translated into a corresponding proof-term. Conservativity ensures the reverse direction: any encoded proposition derivable in the encoded system is provable in the original logic, thereby reducing proof checking to type checking. These two properties are proved individually for the constructive and classical HOL fragments of the modular encoding implemented in the \sys{Lambdapi} standard library in \cite{DBLP:conf/types/Grienenberger22} and together establish the two directions of correctness.

\begin{theorem}[Correctness \cite{DBLP:conf/types/Grienenberger22}] 
Let $\mathcal{L}$ be a language of HOL and $\Delta \vdash_{HOL} \delta$ a proof in ND in HOL. Then, the following holds:
\[
\Delta \vdash_{HOL} \delta 
\quad \Leftrightarrow \quad
(\exists t. \mathcal{HOL}, \|\mathcal{L}\|, \|\Delta\| \vdash t : \| \delta\|).
\]
\end{theorem}

\newpage

\section{Extended Calculus}\label{ap:extendedCalculus}

\subsection{Calculus Rules}

In the rules stated below, we follow the notation given in \cite{LeoIIIthesis}, where a double inference line is used for rules modifying or deleting clauses from the search space. If multiple premises are given, the leftmost premise serves as the \emph{main premise} and denotes the clause being affected. All remaining premises serve as side conditions and remain unchanged. If no conclusion is given, the rule denotes a deleting inference that removes the main premise from the search space.

\begin{mdframed}[
  frametitle={\bfseries Extended Calculus},
  frametitlealignment=\raggedright
]
\begin{minipage}{\textwidth}  
        \begin{minipage}{0.52\linewidth}
        \begin{prooftree}
            \AxiomC{$C \lor [P \, s]^{\nn} \lor [P \, t]^{\pp}$}
            \TightRuleLabel{\RuleDef{LEQ}}
            \UnaryInfC{$C\{\lambda X.s = X/P\} \lor [s \simeq t]^{\pp}$}
        \end{prooftree}
        \end{minipage}
        \hfill
        \begin{minipage}{0.38\linewidth}
            \centering
            \begin{prooftree}
                \AxiomC{$C \lor [P \, s \, s]^{\nn}$}
                \TightRuleLabel{\RuleDef{AEQ}}
                \UnaryInfC{$C\{\lambda X.s = X/P\}$}
            \end{prooftree}
        \end{minipage}
\end{minipage}

\begin{prooftree}
  \AxiomC{$C' \,\lor\, [\,F \, s^{1,1} \, \dots \, s^{1,n} \simeq t^{1}\,]^{\nn} \,\lor\, \cdots \,\lor\, [\,F \, s^{m,1} \, \dots \,s^{m,n} \simeq t^{m}\,]^{\nn}$}
  \TightRuleLabel{\RuleDef{FS}}
  \UnaryInfC{$C\!\left\{\,\lambda X^{1}\ldots X^{n}.\,\varepsilon Z_{\ell}.\,
     \bigwedge_{k=1}^{m}\!\left(\left(\bigwedge_{j=1}^{n} X^{j}=s^{k,j}\right) \Rightarrow Z=t^{k}\right)
     \,/\, F \right\}$}
\end{prooftree}

\raggedright

\begin{minipage}{\linewidth}
    \begin{prooftree}
        \AxiomC{$C \lor [s[\forall X_\tau . u] \simeq t]^{\alpha}$}
        \AxiomC{$v \in Heu^\tau$}
    \RightRightRuleLabel{\RuleDef{HeuInst}}{where $Heu^\tau$ is a set \\of closed terms of type $\tau$}
    \BinaryInfC{$C \lor [s[u\{v/X\}] \simeq t]^{\alpha}$}
    \end{prooftree}
\end{minipage}


\begin{prooftree}
        \AxiomC{$[f X^{1} ... X^{i-1} Y X^{i+1} ... X^n \simeq  f X^{1} ... X^{i-1} Z X^{i+1} ... X^n]^{\nn} \lor [Y \simeq Z]^{\pp}$}
    \TightRuleLabel{\RuleDef{INJ}$^\ddagger$}
    \UnaryInfC{$[f^{inv}X^1 \dots X^{i-1} \; X^{i+1}\dots X^n \; (f \;X^1 \dots X^n) \; \simeq X^i]^{\pp}$}
    \end{prooftree}

\noindent
\begin{minipage}{\linewidth}
    \begin{minipage}{0.4\linewidth}
        \begin{prooftree}
            \AxiomC{$C \lor [s \simeq t]^{\alpha} \lor [s \simeq t]^{\alpha}$}
            \TightRuleLabel{\RuleDef{DD}}
            \doubleLine
            \UnaryInfC{$C \lor [s \simeq t]^{\alpha}$}
            \singleLine
        \end{prooftree}
    \end{minipage}
    \hfill
    \begin{minipage}{0.55\linewidth}
        \begin{prooftree}
            \AxiomC{$C \lor [s[E \; t]]^{\alpha}$}
        \TightRightRuleLabel{\RuleDef{ACI}}{where $E$ is a choice\\operator or $E \in fv(C)$}
        \UnaryInfC{$[t \; x]^{\nn} \lor [t \; (\varepsilon t)]^{\pp}$}
        \end{prooftree}
    \end{minipage}
\end{minipage}

\vspace{1em}

\begin{minipage}{\linewidth}  
        \begin{minipage}{0.47\linewidth}
        \begin{prooftree}
          \AxiomC{$C \lor [s \simeq t]^{\nn}$}
          \AxiomC{$[l \simeq r]^{\pp}$}
          \TightRuleLabel{\RuleDef{PSR}$^\dagger$ }
          \doubleLine
          \BinaryInfC{$C$}
          \singleLine
        \end{prooftree}
        \end{minipage}
        \hfill
        \begin{minipage}{0.47\linewidth}
           \begin{prooftree}
              \AxiomC{$C \lor [s \simeq t]^{\pp}$}
              \AxiomC{$[l \simeq r]^{\nn}$}
              \TightRuleLabel{\RuleDef{NSR}$^\dagger$}
              \doubleLine
              \BinaryInfC{$C$}
              \singleLine
            \end{prooftree}
        \end{minipage}
\end{minipage}
    
\vspace{0.6em}

    \begin{minipage}{\linewidth}  
        \begin{prooftree}
            \AxiomC{$C \lor [s \simeq t]^\alpha$}
            \AxiomC{$[l \simeq r]^{\pp}$}
            \RightRightRuleLabel{\RuleDef{RW}}{if $\exists\sigma$ with $s|_p \equiv l\sigma$ and\\$l\sigma \succ r\sigma$ (w.r.t.\ a term ordering $\succ$)}
            \doubleLine
            \BinaryInfC{$C \lor [s[r\sigma]_p \simeq t]^\alpha$}
            \singleLine
        \end{prooftree}
    \end{minipage}

\begin{prooftree}
    \AxiomC{$[l_1 \simeq r_1]^{\alpha_1} \lor \cdots \lor [l_n \simeq r_n]^{\alpha_n} \quad$}
    \RightLabel{\RuleDef{Simp}}
    \doubleLine
    \UnaryInfC{$[\text{simp}(l_1) \simeq \text{simp}(r_1)]^{\alpha_1} \lor \cdots \lor [\text{simp}(l_n) \simeq \text{simp}(r_n)]^{\alpha_n}$}
    \singleLine
    \vspace*{1mm}
\end{prooftree}

\begin{minipage}{\linewidth}  
        \begin{minipage}{0.47\linewidth}
            \begin{prooftree}
              \AxiomC{$C \lor [s \simeq s]^{\pp}$}
              \RightRuleLabel{\RuleDef{TD1}}{}
              \doubleLine
              \UnaryInfC{}
              \singleLine
            \end{prooftree}
        \end{minipage}
        \hfill
        \begin{minipage}{0.47\linewidth}
            \begin{prooftree}
              \AxiomC{$C \lor [s \simeq t]^{\pp} \lor [s \simeq t]^{\nn}$}
              \RightRuleLabel{\RuleDef{TD2}}{}
              \doubleLine
              \UnaryInfC{}
              \singleLine
            \end{prooftree}
        \end{minipage}
\end{minipage}

\vspace{1em}

\begin{minipage}{\linewidth}  
        \begin{minipage}{0.47\linewidth}
            \begin{prooftree}
              \AxiomC{$C \lor C'$}
              \AxiomC{$D$}
              \RightRightRuleLabel{\RuleDef{CS}}{if $D\sigma \equiv C'$ for \\some substitution $\sigma$}
              \doubleLine
              \BinaryInfC{}
              \singleLine
            \end{prooftree}
        \end{minipage}
        \hfill
        \begin{minipage}{0.47\linewidth}
            \begin{prooftree}
              \AxiomC{$[P\,X]^{\nn} \lor [P(f\,P)]^{\pp}$}
              \RightRuleLabel{\RuleDef{ACD}}{}
              \doubleLine
              \UnaryInfC{}
              \singleLine
            \end{prooftree}
        \end{minipage}
\end{minipage}
    
\vspace{0.8em}
    
\begin{center}
    \footnotesize{$\dagger$: if there exists a substitution $\sigma$ such that 
    $s|_p \equiv l \sigma$ and $l\sigma$ is bigger than $r\sigma$ w.r.t. a term ordering.}\\
      \footnotesize{$\ddagger$: Where $f^{inv}$ is a fresh Skolem constant and $X^1, \dots , X^n$ variables of appropriate type.}
    \end{center}

\end{mdframed}


\subsection{Encoding Details and Status}

Table~\ref{tab:categoriesExt} summarizes the encoding categories of the rules of the extended calculus with respect to the framework introduced in Sec.~\ref{sec:EncDeriv}, and indicates their implementation status. Note that some of the rules (\RuleRef{TD1}, \RuleRef{TD2}, \RuleRef{CS}, and \RuleRef{ACD}) delete clauses from the search space. They never appear in proof certificates and can be omitted during verification. 
\RuleRef{DD} is encoded as a meta-theorem called \lpTerm{delete} which has been added to the \sys{Lambdapi} standard library. 

\begin{table*}[!htbp]
\centering
\caption{Encoding categories for the rules extending the core EP calculus.}
\label{tab:categoriesExt}
\begin{tabular}{l l c@{}}
\toprule
\textbf{Rule} & \textbf{Encoding category} & \textbf{Implemented} \\
\midrule
\RuleRef{TD1}      & \omitRule & - \\
\RuleRef{TD2}      & \omitRule & - \\
\RuleRef{PSR}      & \corrRW & No \\
\RuleRef{NSR}      & \corrRW & No \\
\RuleRef{RW}       & \corrRW & Yes \\
\RuleRef{CS}       & \omitRule & - \\
\RuleRef{LEQ}      & \corrInst & No \\
\RuleRef{AEQ}      & \corrInst & No \\
\RuleRef{ACD}      & \omitRule & - \\
\RuleRef{ACI}      & Instance of a theorem & No \\
\RuleRef{HeuInst} & \FunOnSubst & No\\
\RuleRef{FS}       & \corrInst & No\\
\RuleRef{INJ}      & \FunOnSubst & No \\
\RuleRef{Simp}     & \RwEq & Yes \\
\RuleRef{DD}     & \FunOnClause \, (\lpTerm{delete}) & Yes \\
\bottomrule
\end{tabular}
\end{table*}

\fi

\end{document}